\pdfoutput=1

\documentclass[twocolumn,10pt]{autart}    

\usepackage{graphicx}          
\usepackage{epstopdf}
\usepackage{setspace}                          
\usepackage{amsmath}
\usepackage{enumerate}
\usepackage{graphicx}          
\usepackage{tikz}
\usepackage{cite}
\usepackage{natbib}                          

\usepackage{amsfonts}
\usepackage{algorithm}
\usepackage{algpseudocode}
\usepackage{amssymb}
\usepackage{subfigure}
\usepackage{hyperref}
\hypersetup{
    colorlinks = true,
    citecolor = cyan,
}
\usepackage{wrapfig}

\def\({\left(}
\def\){\right)}
\def\a{\alpha}
\DeclareMathOperator{\tr}{tr}
\newtheorem{assu}{Assumption}[section]
\newtheorem{remark}{Remark}[section]

\newtheorem{example}{Example}

\newtheorem{lemma}{Lemma}[section]
\newtheorem{proposition}{Proposition}
\newtheorem{corollaryx}{Corollary}[section]
\newtheorem{theorem}{Theorem}[section]

\begin{document}
\begin{frontmatter}
\title{Asymptotically Efficient Quasi-Newton Type Identification with Quantized Observations Under Bounded Persistent Excitations\thanksref{footnoteinfo}}
\thanks[footnoteinfo]{The work is supported by National Key R\&D Program of China under Grant 2018YFA0703800, National Natural Science Foundation of China under Grants T2293770, 62025306 and 62303452, China postdoctoral Science Foundation under Grant 2022M720159 and Guozhi Xu Postdoctoral Research Foundation. The material in this paper was not presented at any conference.
Corresponding author:  Yanlong Zhao.}

\author[AMSS]{Ying Wang}\ead{wangying96@amss.ac.cn},
\author[AMSS,UCAS]{Yanlong Zhao}\ead{ylzhao@amss.ac.cn},
\author[AMSS,UCAS]{Ji-Feng Zhang}\ead{jif@iss.ac.cn},
\address[AMSS]{Key Laboratory of Systems and Control, Institute of Systems Science, Academy of Mathematics and Systems Science, Chinese Academy of Sciences, Beijing 100190, P. R. China.}
\address[UCAS]{School of Mathematical Sciences, University of Chinese Academy of Sciences, Beijing 100049, P. R. China.}


\begin{abstract}
This paper is concerned with the optimal identification problem of dynamical systems in which only quantized output observations are available under the assumption of fixed thresholds and bounded persistent excitations. Based on a time-varying projection, a weighted Quasi-Newton type projection (WQNP) algorithm is proposed. With some mild conditions on the weight coefficients, the algorithm is proved to be mean square and almost surely convergent, and the convergence rate can be the reciprocal of the number of observations, which is the same order as the optimal estimate under accurate measurements.
Furthermore, inspired by the structure of the Cram\'er-Rao lower bound, an information-based identification (IBID) algorithm is constructed with adaptive design about weight coefficients of the WQNP algorithm, where the weight coefficients are related to the parameter estimates which leads to the essential difficulty of algorithm analysis. Beyond the convergence properties, this paper demonstrates that the IBID algorithm tends asymptotically to the Cram\'er-Rao lower bound, and hence is asymptotically efficient. Numerical examples are simulated to show the effectiveness of the information-based identification algorithm.

\end{abstract}
\end{frontmatter}

\section{Introduction}

\subsection{Background and Motivations}
Along with the modern science and technology rapid development, quantized systems have been widely applied in practical fields such as industrial systems, networked systems and even biological systems.
For example, i) industrial systems 
\citep{WKS2002,APPCM2018,GMTC2021}:
usually quantized sensors are more cost effective than regular sensors.
In many applications, they are the only ones available during real-time operations.
There are numerous examples of quantized observations such as switching sensors for exhaust gas oxygen, ABS (anti-lock braking systems), and shift-by-wire; photoelectric sensors for positions, and Hall-effect sensors for speed and acceleration for motors; traffic condition indicators in the asynchronous transmission mode networks; and gas content sensors (CO, CO$_2$, H$_2$, etc.) in gas and oil industry.
ii) Networked systems 
\citep{DC2010,SMZ2007}:
thousands, even millions, of sensors are interconnected using a heterogeneous network of wireless systems. On account of limitations of the sensor power or communication bandwidth, the information from each sensor turns to be quantized observations with a finite bit or even 1 bit.
iii) Biological systems \citep{G2003,WZY2003,WYZZ2010}: only two states information, ``excitation" or ``inhibition", are detected from outside of the neuron. When the potential is bigger than the potential threshold, the neuron shows the excitation state, otherwise shows the inhibition state.

Due to the widespread adoption of systems with quantized observations, lots of researches related to the identification of such systems have emerged in the literature 
\citep{WZY2003,WYZZ2010,CGV2011,GGAMW2011,CSM2020,CW2012,WZZG2022,RBH2020,ZZWK2023}.
In addition, numerous methods are proposed to achieve identification with quantized observations such as empirical measure method \citep{WZY2003,WangandYin2007}, expectation maximization method \citep{GGAMW2011}, sign-error type algorithm \citep{CW2012,WZZG2022}, stochastic approximation type algorithm \citep{GZ2013,S2018}, and stochastic gradient type algorithm \citep{GZ2014,ZWZ2021}.
The emergence of these algorithms prompts us to explore how to achieve better identification effect by use of  algorithm designs.
Moreover, the study of the optimal quantized identification algorithms not only could achieve the improvement of identification theory, but also is helpful to improve the resource utilization with the limited communication bandwidth resources in the communication fields.

As known, the unbiased estimator with smaller variance is more efficient than the other in estimation theory. Therefore, the Cram\'er-Rao (CR) lower bound, which is an irreducible lower bound for parameter estimates, comes into our view because it may be used as a criterion to check the effectiveness of a procedure. If the CR lower bound is achieved, then the corresponding estimator is termed efficient.
Therefore, this paper would like to investigate the optimal identification under quantized observations from the point of the CR lower bound.

\subsection{Related literature}
Actually, there are some interesting discussions about the CR lower bound of quantized systems \citep{GZ2014,GK2009,WWY2013}.
For example, \cite{GK2009} and \cite{WWY2013} investigated a detailed study on the CR lower bound and derived its expression under different quantized measurements.
Moreover, some results have also been appeared for asymptotically efficient algorithms under quantized observations in the past two decades 
\citep{WZY2003,WangandYin2007,ZWZ2021,YF2014,You2015,GWYZZ2015,WTZ2018,GD2020}.
\cite{WangandYin2007} proposed a quasi-convex combination estimator (QCCE) that employed empirical measures from multiple sensor thresholds, and established strong consistency and asymptotical optimality of the QCCE under periodic inputs.
The asymptotical efficiency properties of empirical measure method and non-truncated empirical measure method were considered for FIR systems under binary-valued observations and periodic inputs in \cite{WZY2003} and \cite{WTZ2018}, respectively.
Besides, based on empirical measure method, asymptotically efficient algorithms are investigated for the systems with quantized observations and general quantized periodic inputs under various cases  
\citep{GWYZZ2015,GD2020}.
Apart from the off-line algorithms mentioned above, there are also some discussions on online algorithms.
\cite{YF2014} presented a recursive identification method for FIR systems with quantized measurements based on stochastic approximation algorithm with expanding truncation bounds, and proved asymptotically efficient under independent and identically distributed (i.i.d) two-valued random inputs.
\cite{You2015} developed a stochastic approximation type recursive estimator for FIR systems with adaptive binary observations and i.i.d. input signals, and demonstrated it asymptotically approached the CR lower bound.
\cite{ZWZ2021} proposed a stochastic gradient-based recursive algorithm for FIR systems with binary observations, and showed the convergence and asymptotic efficiency under bounded persistent excitations only first-order systems.

 However, almost all of the existing investigations on asymptotically efficient quantized identification algorithms suffer from some fundamental limitations.
 Most of the researches on asymptotically efficient algorithms are based on the empirical measure algorithm, which is off-line and thus is difficult to apply to feedback controls.
 On the other hand, the conditions required are strict in the almost all of the works about online asymptotically efficient algorithms, such as the periodic or two-valued random or i.i.d. inputs, the adaptive and designable thresholds and so on.

 Therefore, the goal of this paper is to develop an asymptotically efficient online algorithm, which could relax or remove the above-mentioned limitations.
  It is our hope that the approach of this paper will open up new avenues for further studies in the area of integrated design of identification and control with quantized constraints.

\subsection{Main contributions}

This paper investigates the asymptotically efficient recursive identification of the systems under quantized observations with multiple thresholds.
The main contributions of this paper can be summarized as follows:

\begin{itemize}
\item  {Inspired by a time-varying projection in \cite{ZZG2022}, a novel weighted Quasi-Newton type projection (WQNP) algorithm is proposed under quantized observations with multiple thresholds.
    With some mild conditions, the WQNP algorithm is proved to be convergent in both mean square and almost sure sense under bounded persistent excitations with the help of an inner product type Lyapunov function.
    Besides, the convergence rate can achieve the reciprocal of the number of observations under a proper requirement of weight coefficients, which is the same order as that under accurate measurements.}

\vspace{0.15cm}
\item {This paper gives the CR lower bound of the system under multiple-threshold quantized observations. Then, based on the recursive form of its CR lower bound to design the weight coefficients of the WQNP algorithm, an information-based identification (IBID) algorithm is constructed, whose adaptive weight coefficients depend on the parameter estimates.
    Besides, the convergence rate is proved that can reach the reciprocal of the time step by combining the inner product type and cross product type Lyapunov function methods.
    Moreover, the IBID algorithm is asymptotically efficient under bounded persistent excitations.
    In contrast with \cite{WangandYin2007}, the algorithm is an asymptotically efficient online algorithm under non-periodic or non-independent signals. }

\vspace{0.15cm}
\item { The theoretical analysis method is different from the existing quantized identification algorithms.
    This paper adopts an idea of higher moment acceleration to solute the strong coupling between the weighted coefficients and the estimates of the IBID algorithm in the cross product type Lyapunov function method.
    It is worth mentioning that Markov inequality and the higher moments of estimation errors are used to establish the convergence rate of the cross product type Lyapunov function.}
\end{itemize}

The rest of this paper is organized as follows. Section \ref{sec:PF} describes the identification problem under multiple sensor thresholds. Section \ref{sec:WQNPA} presents the WQNP algorithm, and demonstrates its convergence properties. Section \ref{sec:efficiency} constructs the IBID algorithm based on the CR lower bound, and establishes its convergence properties and asymptotic efficiency. All of the proofs of the main results are uniformly provided in Section \ref{sec:proof}.
Section \ref{sec:sim} supplies a numerical example to show the main results. Section \ref{sec:con} gives the concluding remarks and related future works.

\noindent\emph{Notation.} In this paper,
$\mathbb{R}^n$ and $\mathbb{R}^{n\times n}$ are the sets of $n$-dimensional real vectors and $n\times n$ dimensional real matrices, respectively. $I_n$ is an $n$-dimension identity matrix. $\|\cdot\|$ is the Euclidean norm, i.e, $\|x\|=\(\sum_{i=1}^nx_i^2\)^{\frac{1}{2}}$ for the vector $x\in\mathbb{R}^n$ and $\|A\|=\sqrt{\(\lambda_{\max}(AA^T)\)}$ for the matrix $A\in\mathbb{R}^{n\times n}$. Besides, the trace of the matrix $A$ is $\tr(A)=\sum_{i=1}^na_{ii}$.  For the matrix $A_k$ , denote $A_k=O\(\frac{1}{k}\)$ as $\|A_k\|=O\(\frac{1}{k}\)$ and $A_k=o\(\frac{1}{k}\)$ as $\|A_k\|=o\(\frac{1}{k}\)$. The function $I_{\{\cdot\}}$ denotes the indicator function, whose value is 1 if its argument (a formula) is true, and 0, otherwise.

\section{Problem formulation}\label{sec:PF}
\subsection{Observation model}

Consider the following dynamic linear system
\vspace{-5pt}
\begin{gather}\label{M}
y_{k}= \phi^T_{k}\theta +d_{k},\quad k=1,2,\ldots,
\end{gather}
where $k$ is the time index, $\phi_{k}\in\mathbb{R}^n$, $\theta\in\mathbb{R}^n$, and $d_{k}\in\mathbb{R}$ are the regressor, unknown but constant parameter vector, and noise at time $k$, respectively. The system output $y_{k}$ is measured by a sensor of $m$ thresholds $-\infty<C_1<C_2<\cdots<C_m<\infty$. The sensor is represented by a set of $m$ indicator function, which is given by
\begin{equation}\label{qk-mul}
q_k
=\left\{
   \begin{array}{ll}
     0, & \hbox{if } ~ y_k \leq C_1; \\
     1, & \hbox{if } ~ C_{1}<y_k\leq C_2;\\
     \vdots & \hbox{} \vdots \\
     m, & \hbox{if } ~ y_k>C_m;
   \end{array}
 \right.
\end{equation}
which can also be represented as $q_k=\sum_{i=0}^{m} i I_{\{C_{i} < y_k \leq C_{i+1}\}}$, where $C_0=-\infty$ and $C_{m+1}=\infty$.

\subsection{Assumptions}

In order to proceed our analysis, we introduce some assumptions concerning priori information of the unknown parameter, the regressors and the noises.

\begin{assu}\label{AP}The prior information on the unknown parameter $\theta$ is that $\theta\in \Omega \subset \mathbb{R}^n$ with $\Omega$ being a bounded convex set. And denote $\bar{\theta}=\sup_{\eta\in\Omega}\|\eta\|.
$\end{assu}

\begin{assu}\label{AI}
The vector sequence $\{\phi_{k}\}$ is supposed to be bounded persistently exciting, i.e.,
\begin{equation}\label{def:pe}
{\liminf_{k\rightarrow\infty}\frac{1}{k}\sum_{l=1}^{k} \phi_{l}\phi_{l}^T>0},
\end{equation}
and $\sup_k\|\phi_{k}\|\leq \bar{\phi}<\infty$.
\end{assu}

\begin{assu}\label{AN}
Assume that $\{d_{k}\}$ is a sequence of independent and identically normally distributed variables following $N(0,\sigma^2)$. The distribution and density functions of $d_1$ are denoted as $F(\cdot)$ and $f(\cdot)$, respectively.
\end{assu}

\begin{remark}\label{nA1}
Actually, the median $\mu$ of the noise could be estimated similarly to \cite{WZZG2022} when $\mu\neq0$. Therefore, without loss of generality, we assume that $\mu=0$ throughout the paper.
{Furthermore, the noise under Assumption \ref{AN} can be generalized to the one that the second derivation of the logarithm density function is less than zero (i.e., $\frac{d^2\ln f(x)}{d x^2}<0$), and the density function of the noise satisfies $\min_{1\leq i\leq m \atop x\in[C_i-\bar{\phi}\bar{\theta} ,C_i+\bar{\phi}\bar{\theta}]}f(x)>0$.}
\end{remark}

The goal of this paper is to develop an online asymptotically efficient algorithm to estimate the unknown parameter $\theta$  based on the information from input $\phi_{k}$, quantized observation $q_{k}$, and the stochastic property of system noise $d_{k}$ under bounded persistent excitations.

\section{The WQNP algorithm}\label{sec:WQNPA}
This section will construct a Quasi-Newton type identification algorithm under quantized observations, and establish its convergence properties.

\subsection{Algorithm design}
For the simplicity of description, denote
$F_i(x)=F(C_i-x)$, $f_i(x)=f(C_i-x)$, for $i=0,\ldots,m+1$, and $H_i(x)=F_i(x)-F_{i-1}(x)$, $h_i(x)=f_i(x)-f_{i-1}(x)$ for $i=1,\ldots,m+1$. Moreover, denote
\begin{align}\label{realFf}
F_{i,k}=F_i\(\phi^T_k\theta\),\hbox{  } f_{i,k}=f_i(\phi^T_k\theta),
\end{align}
and their estimates based on $\hat\theta_{k-1}$ as
\begin{align}\label{estFf}
\hat{F}_{i,k}=F_i(\phi^T_k\hat\theta_{k-1}),
\hbox{ }
\hat{f}_{i,k}=f_i(\phi^T_k\hat\theta_{k-1}),
\end{align}
 for $i=0, \ldots, m+1$.
Correspondingly, denote
\begin{align}\label{realHh}
H_{i,k} = H_{i}(\phi^T_k\theta), \hbox{ }
h_{i,k}= h_{i}(\phi^T_k\theta),
\end{align}
and their estimates as
\begin{align}\label{estHh}
\hat{H}_{i,k}=H_{i}(\phi^T_k\hat\theta_{k-1})\hbox{  } \hat{h}_{i,k}=h_{i}(\phi^T_k\hat\theta_{k-1}),
\end{align}
for $i=1, \ldots, m+1$. Hence,
$\mathbb{E }q_k =\sum_{i=1}^{m+1}(i-1){H}_{i,k}.$

Next, we would like to introduce the idea of the Quasi-Newton type identification algorithm under quantized observations.
Actually, the identification problem of unknown parameter $\theta$ is to find the roots of
$$u_k(\hat \theta)=\sum_{i=1}^{m+1}(i-1){H}_{i,k}
-\sum_{i=1}^{m+1}(i-1){H}_{i}(\phi_k^T\hat\theta),$$
for all $k\geq 0$.
Note $\sum_{i=1}^{m+1}(i-1){H}_{i,k}$ is unavailable due to the existence of unknown parameter $\theta$, and $q_k$ is available with its expectation $\sum_{i=1}^{m+1}(i-1){H}_{i,k}$. Therefore, we
replaced $\sum_{i=1}^{m+1}(i-1){H}_{i,k}$ with $q_k$ in $u_k(\hat \theta)$.
By instrumental variable method \citep{S_LS1983}, we use $\phi_k$-s instrumental variable to define the vector-valued scores
\vspace{-5pt}
\begin{align}\label{U_k}
 U_k(\hat\theta)&=-\sum_{l=1}^k\(q_l
-\sum_{i=1}^{m+1}(i-1){H}_{i}(\phi_l^T\hat\theta)\)\phi_l.
\end{align}
whose Jacobian matrix is used to construct the Newton-type step.
Then, we calculate $\frac{\partial U_k(\hat\theta)}{\partial\hat\theta}$ as
\vspace{-5pt}
\begin{align}\label{U_k}
 \frac{\partial U_k(\hat\theta)}{\partial\hat\theta}
 =-\sum_{l=1}^k\sum_{i=1}^{m+1}(i-1){h}_{i}(\phi_l^T\hat\theta)\phi_l\phi_l^T.
\end{align}
We generalize the above calculated Newton step as
\vspace{-5pt}
\begin{align}\label{P_k0}
 P_k =\sum_{l=1}^k\beta_l\phi_l\phi_l^T.
\end{align}
Then, based on the idea of recursive least squares, we construct the identification algorithm as
\vspace{-5pt}
\begin{align}
\hat\theta_{k}&=\hat\theta_{k-1}+
a_kP_{k-1}\phi_{k}\(q_k-\sum_{i=1}^{m+1}(i-1)\hat{H}_{i,k}\),\nonumber\\
a_k&=\frac{1}{1+\beta_k\phi_{k}^TP_{k-1}\phi_{k}},\nonumber\\
P_{k}&=P_{k-1}-a_k\beta_kP_{k-1}\phi_{k}\phi_{k}^TP_{k-1}.\nonumber
\end{align}
Then, we design the weight coefficients $\alpha_{i,k}$ on the quantized observation $q_k$ to adjust the performance of the identification algorithm, i.e.,
\vspace{-8pt}
$$s_k=\sum_{i=1}^{m+1} \alpha_{i,k} I_{ \{ C_{i-1}< y_k \leq C_{i}\}}.$$
Moreover, we utilize the specific time-varying projection operator in \cite{ZZG2022} to guarantee the boundness of estimates, which is also helpful in the convergence analysis based on Lyapunov function method.
Based on the above idea, a weighted Quasi-Newton type projection (WQNP) algorithm is constructed as Algorithm \ref{Algorithm1}.
\vspace{-5pt}
\begin{algorithm}
\caption{The WQNP Algorithm}
\label{Algorithm1}
Beginning with an initial values $\hat\theta_0\in \Omega$ and an positive definitive matrix ${P}_0\in\mathbb{R}^{n\times n}$, the algorithm is recursively defined at any $k\geq0$ as follows:
\begin{algorithmic}[1]
\State  {Weighted conversion of the quantized observations:
\vspace{-5pt}
\begin{equation}\label{sk-mul}
s_k=\sum_{i=1}^{m+1} \alpha_{i,k} I_{ \{ C_{i-1}< y_k \leq C_{i}\}}.
\end{equation}}
\vspace{-5pt}
\State Estimation:
\vspace{-5pt}
\begin{align}
\label{update}\hat\theta_{k}&=\Pi_{P_k^{-1}}\(\hat\theta_{k-1}+
a_kP_{k-1}\phi_{k}\tilde{s}_k\),\\
\label{tildesk}\tilde{s}_k&=s_k-\sum_{i=1}^{m+1} \alpha_{i,k}\hat{H}_{i,k},\\
\label{ak}a_k&=\frac{1}{1+\beta_k\phi_{k}^TP_{k-1}\phi_{k}},\\
\label{variance}P_{k}&=P_{k-1}-a_k\beta_kP_{k-1}\phi_{k}\phi_{k}^TP_{k-1},
\end{align}
where $\hat{H}_{i,k}$ are defined in  (\ref{estHh}). Besides, $\Pi_{Q}(\cdot)$ is the projection mapping defined as
\vspace{-5pt}
\begin{gather}\label{projection}
\Pi_{Q}(x)= \arg\min_{z\in\Omega}\|x-z\|_Q, \forall x\in\mathbb{R}^n,
\end{gather}
where $\|\cdot\|_Q$ is defined as $\|\eta\|_Q=\sqrt{\eta^TQ\eta}, \forall \eta\in\mathbb{R}^n$ and Q is a positive definitive matrix.
\end{algorithmic}
\end{algorithm}


\begin{remark}
It is worth noticing that when the quantized output is binary-valued observation (i.e., $m=1$) and the dimension of the unknown parameter $\theta$ is one (i.e., $n=1$), the WQNP algorithm can degrade into the unified stochastic gradient-based recursive algorithm in \cite{ZWZ2021}. More specifically, the innovation of the quantized observation in (\ref{tildesk}) can be rewritten as $\tilde s_{k}= (\alpha_{2,k}-\alpha_{1,k})\(F(C_1-\phi^T_k\hat\theta_{k-1}) -I_{\{y\leq C_1\}}\).$
Therefore, the WQNP algorithm is a general extension of the algorithm in \cite{ZWZ2021} from binary observations to multiple sensor threshold observations.
\end{remark}

\subsection{Convergence properties}
Before establishing the convergence, the following assumption about the weight coefficient is given.
\begin{assu}\label{AAB}
 The weight coefficients $\alpha_{i,k}$ $(i=1, \ldots, m+1)$ and $\beta_k$ are scalars satisfying $ - \infty< \underline \alpha\leq \alpha_{1,k} < \alpha_{2,k} < \cdots <\alpha_{m+1,k} \leq \overline\alpha< \infty$ with $\alpha_{m+1,k} -\alpha_{1,k} \geq \alpha >0$ and $0<\underline\beta\leq \beta_k\leq\overline\beta<\infty $, respectively.
 Besides, the weight coefficients satisfy $\frac{2\alpha}{\overline{\beta}}\cdot \min_{x\in[C_i-\bar\phi\bar\theta,C_i+\bar\phi\bar\theta]\atop 1\leq i\leq m} f(x)>1-\frac{1}{n}$.
\end{assu}
\begin{theorem}\label{thm:cov}
If Assumptions \ref{AP}-\ref{AN} and \ref{AAB} hold, then the WQNP algorithm is convergent both in mean square and high rank square, i.e.,
\vspace{-5pt}
\begin{gather}\label{mc}
\lim_{k\to \infty}
\mathbb{E}\tilde\theta_{k}^T\tilde\theta_{k}=0 \quad \hbox{and} \quad
\lim_{k\to \infty} \mathbb{E}\|\tilde\theta_{k}\|^{2r}=0, \quad
\end{gather}
and there exists a positive real number $\nu<\infty$ such that
\vspace{-5pt}
\begin{gather}\label{mcr}
\mathbb{E}\|\tilde\theta_{k}\|^{2}=O\left(\frac{1}{k^{\nu}}\right) \hbox{ and   }  \mathbb{E}\|\tilde\theta_{k}\|^{2r}=O\left(\frac{1}{k^{r\nu}}\right),
\end{gather}
for $r=2,3,\ldots$ Besides, the WQNP algorithm is also convergent almost surely, i.e.,
\vspace{-5pt}
\begin{gather}
\lim_{k\to \infty}\tilde\theta_{k}=0 \quad \hbox{a.s.},\nonumber
\end{gather}
where $\tilde\theta_{k}=\hat\theta_{k}-\theta$ is the estimation error.
\end{theorem}
 The proof of Theorem \ref{thm:cov} is supplied in Section \ref{proof:thm:cov}.

\begin{remark}
{ Theorem \ref{thm:cov} establishes the convergence properties of the WQNP algorithm for high-order parameter systems with  quantized observations while \cite{ZWZ2021} shows the convergence properties for 1-order parameter systems.
The key difficulty of the proof is how to guarantee the compression coefficient is less than 1, which is related to dealing with the non-commutative matrices. Two techniques are applied in this part. First, a time-varying projection operator is introduced to deal with the product of the non-commutative matrix $P_k\phi_k\phi_k^T$ and keep the boundness of estimates. Besides, the boundness of estimates and regressor is used to ensure $\underline f=\min_{1\leq i\leq m}
\min_{x\in[C_i-\bar{\phi}\bar{\theta},C_i+\bar{\phi}\bar{\theta}]}f(x)>0$, and then make the compression factor $1-\frac{2\a\underline f}{\bar{\beta}}$ in (\ref{kk-hh}) less than 1.
}
\end{remark}

Besides the convergence, the convergence rate is another major problem that should be made clear.

\begin{theorem}\label{thm:covrok}
{ Under the conditions of Theorem \ref{thm:cov}, if the condition (\ref{def:pe}) in Assumption \ref{AI} is enhanced as there exist an positive integer $h$ and positive number $\delta>0$ such that
$\frac{1}{h}\sum_{l=k+1}^{k+h} \phi_{l} \phi_{l}^T\geq\delta^2I_n$},
 and
 \vspace{-5pt}
\begin{align}\label{abc}
\frac{\alpha}{\overline{\beta}}>\(
2\inf_{k}\min_{1\leq i\leq m}\min_{\vartheta\in\Omega}f(C_i-\phi_k^T\vartheta)\)^{-1},
\end{align}
then the WQNP algorithm has a mean square convergence rate as $O\(\frac{1}{k}\)$, i.e.,
\vspace{-5pt}
\begin{gather}
\mathbb{E}\|\tilde\theta_{k}\|^2= O\(\frac{1}{k}\),\nonumber
\end{gather}
\vspace{-5pt}
{where $\alpha$ and $\overline{\beta}$ are defined in Assumption \ref{AAB}. }
\end{theorem}
The proof of Theorem \ref{thm:covrok} is put in Section \ref{proof:thm:covrok}.

\begin{remark}
 Theorem \ref{thm:covrok} describes the fact that even under quantized observations, the convergence rate of $O\(\frac{1}{k}\)$ can be achieved with a suitable design of weight coefficients in the WQNP algorithm (\ref{update})-(\ref{variance}), which is the same rate as the case with accurate measurements.
 \end{remark}

 Similar to the proof of Theorem \ref{thm:cov}, the following corollary can be derived directly, which is concerned with high rank square convergence rate.

 \begin{corollaryx}\label{thm:covrok2} Under the condition of Theorem \ref{thm:covrok}, we have $\mathbb{E}\|\tilde\theta_{k}\|^{2r}= o\(\frac{1}{k}\)$ for $r=2,3,\ldots$
\end{corollaryx}

\section{Asymptotically efficient algorithm}\label{sec:efficiency}
This section focuses on how to design and analyse the optimal identification algorithm under quantized observations. To realize it, we give a criterion, the CR lower bound under quantized observations, based on which an asymptotically efficient algorithm is constructed.

\subsection{Cram\'er-Rao lower bound}
Aiming at the system (\ref{M}) with quantized observations (\ref{qk-mul}), the following proposition establishes the CR lower bound of parameter estimates.
\begin{proposition}\label{thm:CRLB}
For the system (\ref{M}) with quantized observations (\ref{qk-mul}), the CR lower bound is
\vspace{-5pt}
\begin{gather}\label{crlb}
\Delta_{k}=\(\sum_{l=1}^k\rho_l\phi_l\phi_l^T\)^{-1},
\end{gather}
\vspace{-5pt}
where
\vspace{-5pt}
\begin{gather}\label{rhol}
\rho_l=\sum_{i=1}^{m+1}\frac{h_{i,l}^2}{H_{i,l}},
\end{gather}
\vspace{-5pt}
with $h_{i,l}$ and $H_{i,l}$ defined in (\ref{realHh}) for $i=1,\ldots,m+1$.
\end{proposition}
The proof of Proposition \ref{thm:CRLB} is supplied in Section \ref{proof:thm:CRLB}.
To understand the relationship between identification under quantized observations and the one under accurate observations, the following proposition is given.

\begin{proposition}\label{thm:lambdaq}
Under Assumption \ref{AN}, $\rho_l$ defined in (\ref{rhol}) satisfies
$\lim_{\max_{i=1, \ldots, m+1}(C_i - C_{i-1}) \to 0}\rho_l = \frac{1}{\sigma^2}$
where $\sigma^2$ is the covariance of $d_l$.
\end{proposition}
The proof of Proposition \ref{thm:lambdaq} is supplied in Section \ref{proof:thm:lambdaq}.
\begin{remark}
 The CR lower bound of the system (\ref{M}) with accurate observations is $\(\frac{1}{\sigma^2}\sum_{l=1}^k\phi_l\phi_l^T\)^{-1}$.
 Combined it with Proposition \ref{thm:lambdaq}, we find that the influence of quantized observations on the identification effect can be represented by the CR lower bound to some extent.
 \end{remark}

\subsection{The IBID algorithm}
This part will construct an asymptotically efficient algorithm with a proper design of weight coefficients on the WQNP algorithm, which is based on CR lower bound.

By the structure of CR lower bound, $\Delta_k$ defined in (\ref{crlb}) can be written
recursively as
\begin{align}\label{rec-sig}
\Delta_k= \Delta_{k-1}-\frac{\rho_k\Delta_{k-1} \phi_k \phi_k^T \Delta_{k-1}}{1+\rho_k\phi^T_k \Delta_{k-1}\phi_k}.
\end{align}
Since $\rho_k$ depends on the unknown parameter $\theta$, we estimate it by use of $\hat\theta_{k-1}$ as
$\hat\rho_k=\sum_{i=1}^{m+1}\frac{\hat{h}^2_{i,l}}{\hat{H}_{i,l}}$,
where $\hat{H}_{i,l}$ and $\hat{h}_{i,l}$ are defined in (\ref{estHh}).

Note that $P_k$ in recursive least square algorithm  could represent the covariance of estimation error in to some extent, which enlightens us to design its
weight coefficient as the estimate of CR lower bound coefficients, i.e.,
\begin{align}\label{def:hatbeta}
\beta_k=\hat\rho_k=\sum_{i=1}^{m+1}\frac{\hat{h}^2_{i,l}}
{\hat{H}_{i,l}}\triangleq\hat\beta_k.
\end{align}
Moreover, noticing (\ref{U_k}) and (\ref{P_k0})  during the structure process of Newton step, we have
$\beta_k=-\sum_{i=1}^{m}\alpha_{i,k}\hat{h}_{i,k}$.
Therefore, the weight coefficient of the weighted conversion is designed as
\begin{align}\label{opt:alpha}
\alpha_{i,k}=\hat\a_{i,k}\triangleq-\frac{\hat{h}_{i,k}}{\hat{H}_{i,k}},\quad i=1,\ldots,m+1.
\end{align}
From Lemma \ref{lem:dechoH} in Section \ref{proof:thm:condlamb} and the boundness of the estimate $\hat{\theta}_k$ and the regressor $\phi_k$, the following proposition can be established directly to illustrate the properties of $\hat\alpha_{i,k}(i=1,\ldots,m+1)$ and $\hat\beta_k$.

\begin{proposition}\label{thm:condlamb}
Denote
$$\hat\a=\inf_k\min_{x \in \Omega} \(\frac{{f}_{m+1}(x)}{1- {F}_{m+1}(x)} + \frac{{f}_{1}(x)}{{F}_{1}(x)}\).$$
Then, $\hat\beta_k$ and $\hat\alpha_{i,k}$ defined by (\ref{def:hatbeta})-(\ref{opt:alpha}) satisfy $0<\hat\beta_k<\infty$, $-\infty<\hat\a_{1,k} <\cdots<\hat\a_{m+1,k}<\infty$ and $\hat\a_{m+1,k}-\hat\a_{1,k} \geq \hat\a>0$.
\end{proposition}

Based on the WQNP algorithm and the weight coefficients in (\ref{def:hatbeta})-(\ref{opt:alpha}), an IBID algorithm is constructed as Algorithm \ref{Algorithm2}.

\begin{algorithm}
\caption{The IBID Algorithm}
\label{Algorithm2}
Beginning with an initial values $\hat\theta_0\in \Omega$ and an positive definitive matrix $\hat{P}_0\in\mathbb{R}^{n\times n}$, the algorithm is recursively defined at any $k\geq0$ as follows:
\begin{algorithmic}[1]
\State Update of the adaptive weight coefficients:
\vspace{-5pt}
\begin{align}\label{opt:ab}
\hat\a_{i,k}=-\frac{\hat{h}_{i,k}}{\hat{H}_{i,k}}
\quad \hbox{ and } \quad
\hat\beta_k=\sum_{i=1}^{m+1}\frac{\hat{h}^2_{i,k}}{\hat{H}_{i,k}},
\end{align}
where $\hat{h}_{i,k}$ and $\hat{H}_{i,k}$ are defined as (\ref{estHh}).
\State { Weighted conversion of the quantized observations:
\vspace{-5pt}
\begin{equation}\label{sk-mul}
s_k=\sum_{i=1}^{m+1} \hat\alpha_{i,k} I_{ \{ C_{i-1}< y_k \leq C_{i}\}}.
\end{equation}}
\State Estimation:
\vspace{-5pt}
\begin{align}
\label{opt:update}\hat\theta_{k}&= \Pi_{\hat{P}_{k}^{-1}}
\(\hat\theta_{k-1}+\hat{a}_k\hat{P}_{k-1}\phi_{k}\tilde{s}_k\),\\
\label{opt:tildesk}\tilde{s}_k&= s_k-\sum_{i=1}^{m+1} \hat\a_{i,k}\hat{H}_{i,k}, \\
\label{opt:a_k}\hat{a}_k&=\frac{1}{1+\hat{\beta}_k\phi_{k}^T\hat{P}_{k-1}\phi_{k}},\\
\label{opt:variance}\hat{P}_{k}&=\hat{P}_{k-1}
-\hat{a}_k\hat{\beta}_k\hat{P}_{k-1}\phi_{k}^T\phi_{k}\hat{P}_{k-1}.
\end{align}
\end{algorithmic}
\end{algorithm}

\begin{remark}
Different from the WQNP algorithm and the weighted least square algorithm, the weight coefficients $\alpha_{i,k}$ and $\beta_k$ of the IBID algorithm are related to the estimates. This leads to the essential difficulty of algorithm analysis since the properties of the adaptive weight coefficients and the convergence of the estimate are interdependent, which make the inner product type Lyapunov function method no longer applicable. Therefore, we introduce the cross product type Lyapunov function method to analyze the convergence rate of the IBID algorithm.
\end{remark}

\subsection{Convergence properties}
The following theorem shows the convergence and the optimal convergence rate of the IBID algorithm.
\begin{theorem}\label{thm:opt-cov}
If Assumptions \ref{AP}-\ref{AN} hold and the noise density function satisfies
\begin{gather}\label{ANN}
2\min_{x\in[C_i-\bar\phi\bar\theta,C_i+\bar\phi\bar\theta]}f(x) \geq\max_{x\in[C_i-\bar\phi\bar\theta,C_i+\bar\phi\bar\theta]}f(x),
\end{gather}
for $i=1,\ldots,m$, then the IBID algorithm is convergent in both mean square and almost sure sense, i.e.,
$\lim_{k\to \infty}
\mathbb{E}\tilde\theta_{k}^T\tilde\theta_{k}=0$ and
$\lim_{k\to \infty}\tilde\theta_{k}=0, a.s.$
Besides, the mean square convergence rate is
\vspace{-10pt}
$$\mathbb{E}\|\tilde\theta_ {k}\|^{2}=O\left(\frac{1}{k}\right).$$
\vspace{-5pt}
\end{theorem}
\vspace{-10pt}
The proof of Theorem \ref{thm:opt-cov} is supplied in Section \ref{proof:thm:opt-cov}.
\begin{remark}
The noise condition (\ref{ANN}) is mainly used to guarantee that the convergence of the inner product type
Lyapunov function. This keeps in essence $\frac{f(C_i-\phi^T_ {k}\grave{\theta}_{i,k-1})} {f(C_i-\phi^T_ {k}\hat{\theta}_{i,k-1})}>\frac{1}{2}$ in (\ref{lambda_k_u})  hold for $i=1,\cdots, m$, where  $\grave{\theta}_{i,k-1}$ with $\phi^T_{k}\grave{\theta}_{i,k-1}$ in the interval between $\phi^T_{k}\theta$ and $\phi^T_{k}\hat\theta_{i,k-1}$.
This point is also the key difficulty in the convergence analysis of the IBID algorithm. This will be left as an open question. An possibly effective way in the authors' view is removing the limitation of the projection and using the covariance matrix of estimation error to analyze the convergence analysis of the IBID algorithm.
\end{remark}

According to the proof of Theorems \ref{thm:cov} and \ref{thm:opt-cov}, the following corollary is derived directly, which is on the high-rank square convergence rate of the IBID algorithm.

\begin{corollaryx} \label{opt:hmcr}
Under the condition of Theorem \ref{thm:opt-cov}, the IBID algorithm is convergent in high rank square with $\mathbb{E}\|\tilde{\theta}_k\|^{2r}=o\(\frac{1}{k}\)$, for $r=2,3,\ldots$
\end{corollaryx}

\subsection{Asymptotical efficiency}
The following theorem shows $\hat{P}_k$ of IBID algorithm represents the covariance of estimation error to some extent.

\begin{theorem}\label{thm:pktocr}
If Assumptions \ref{AP}-\ref{AN} and (\ref{ANN}) hold, then $\hat{P}_k$ defined in (\ref{opt:variance}) has the following property,
\begin{align}
\lim_{k\to\infty}k(\mathbb{E}\hat{P}_k-\Delta_k)=0.\nonumber
\end{align}
\end{theorem}
The proof of Theorem \ref{thm:pktocr} is supplied in Section \ref{proof:thm:pktocr}.
The following theorem demonstrates that the IBID algorithm can achieve the CR lower bound asymptotically, which implies that the IBID algorithm is asymptotically efficient and optimal.

\begin{theorem}\label{thm:ae}
If Assumptions \ref{AP}-\ref{AN} and (\ref{ANN}) hold, then the IBID algorithm is asymptotically efficient, i.e.,
\begin{gather}
\lim_{k\to \infty} k\(\mathbb{E}\tilde\theta_{k}\tilde\theta_k^T -\Delta_k\)=0.\nonumber
\end{gather}
\end{theorem}
\vspace{-10pt}
The  proof of Theorem \ref{thm:ae} is put in Section \ref{proof:thm:ae}.

\section{Proofs of the main results}\label{sec:proof}
\subsection{Proof of Theorem \ref{thm:cov}}\label{proof:thm:cov}
Before proving the convergence of the WQNP algorithm,  some lemmas are collected and established, which are frequently used in the analysis of convergence.
\begin{lemma}[\cite{CM1987}]\label{lem:proj}
For the bounded convex set $\Omega$, the projection is defined as $\Pi_{Q}(x)= \arg\min_{z\in\Omega}||x-z||_Q$
 for all $x\in\mathbb{R}^n$, where $Q$ is a positive definitive matrix.  Then, for all $x\in
\mathbb{R}$ and $x^*\in \Omega$, it holds
$\left\|\Pi_{Q}(x)-x^*\right\|_Q\leq \left\|x-x^*\right\|_Q$.
\end{lemma}
\vspace{-5pt}
\begin{lemma}[\cite{ZZG2022}]\label{lemma4_ZZG}
Let $X_1$, $X_2, \cdots$ be any bounded sequence of vectors in $\mathbb{R}^n(n\geq 1)$. Denote $A_k=A_0+\sum_{i=1}^{k}X_iX^T_i$ with $A_0>0$. Then, it holds that $\sum_{k=1}^{\infty}(X_i^TA_k^{-1}X_i)^2<\infty$.
\end{lemma}
\vspace{-10pt}
\begin{lemma}[\cite{Chen2002}] \label{Lemma 1.2.2_Chen2002}
 Let $(v_k,\mathcal{F}_k)$, $(w_k,\mathcal{F}_k)$ be two nonnegative adapted sequences.
If $\mathbb{E}(v_{k+1}|\mathcal{F}_k)\leq v_k+w_k$ and $\mathbb{E}\sum_{k=1}^{\infty}w_{k}<\infty$, then $v_k$ converges a.s. to a finite limit.
\end{lemma}
\vspace{-5pt}
\begin{lemma}[\cite{ZWZ2021}] \label{Lemma1_ZWZ2021}
For any given positive integer $l$ and $a, b\in \mathbb{R}$, the following results hold
\vspace{-8pt}
\begin{align}
&\prod_{i=l+1}^k\left(1-\frac{a}{i}\right)
=O\left(\(\frac{l}{k}\)^a\right),\nonumber\\
&\sum_{l=1}^{k}\prod_{i=l+1}^k\left(1-\frac{a}{i}\right)
\frac{1}{l^{1+b}}=
\left\{
\begin{aligned}
&O\left(\frac{1}{k^a}\right),a<b,\\
&O\left(\frac{\ln k}{k^a}\right),a=b,\\
&O\left(\frac{1}{k^b}\right),a>b.
\end{aligned}
\right.\nonumber
\end{align}
\end{lemma}
\vspace{-5pt}
\begin{lemma}\label{prop:pk}
Under Assumption \ref{AI}, $P_k$ defined in (\ref{variance}) has the following properties: i) the inverse of $P_k$ follows $P_{k}^{-1}=P_{k-1}^{-1}+\beta_{k}\phi_{k}\phi_{k}^{T}$; ii) For any initial $P_0>0$,
\begin{equation}\label{Pkok}
0 \leq P_k\leq P_{k-1} \quad \hbox{and} \quad P_k=O\(\frac{1}{k}\).\nonumber
\end{equation}
\end{lemma}
\vspace{-10pt}
{\bf Proof:}  From (\ref{variance}), we have
$P_{k}^{-1}=P_{k-1}^{-1}+\beta_{k}\phi_{k}\phi_{k}^{T}.$
Then, by $P_{k-1}P_{k}^{-1}=I_n+P_{k-1}\beta_{k}\phi_{k}\phi_{k}^{T}$ and iterating the right parts of last equation,
one can get
$P_{k}^{-1}=P_{0}^{-1}+\sum_{i=1}^{k}\beta_{i}\phi_{i}\phi_{i}^{T},$
Consequently, by $\beta_i \geq \underline{\beta}>0$ and
Assumption \ref{AI}, the conclusion is true.
\qed

\begin{lemma} \label{projerror}
If Assumptions \ref{AI} and \ref{AAB} hold, then
\vspace{-5pt}
\begin{equation}
\left\|\tilde{\theta}_{k+j}-\tilde{\theta}_{k}\right\|\leq
j(m+1)|\bar{\alpha}|\bar{\phi}\|P_k\|, j\geq 0 \nonumber.
\end{equation}
\end{lemma}
\vspace{-10pt}
{\bf Proof:}
If $j=0$, then the conclusion is true. Otherwise,
\vspace{-10pt}
\begin{align}\label{pe1}
\|\tilde{\theta}_{k+j}-\tilde{\theta}_{k}\|
=\|\hat{\theta}_{k+j}-\hat{\theta}_{k}\|
\leq\sum_{l=k+1}^{k+j}\|\hat{\theta}_{l}-\hat{\theta}_{l-1}\|.\
\end{align}
\vspace{-5pt}
By Lemma \ref{lem:proj} and (\ref{update}), we have
\vspace{-5pt}
\begin{align}
&\|\hat{\theta}_{l}-\hat{\theta}_{l-1}\|_{P_{l}^{-1}}^2
=\|\Pi_{P_l^{-1}}\(\hat\theta_{l-1}+a_lP_{l-1}\phi_{l}\tilde{s}_l\)
-\hat{\theta}_{l-1}\|_{P_{l}^{-1}}^2\nonumber\\
&\leq \|a_lP_{l-1}\phi_{l}\tilde{s}_l\|_{P_{l}^{-1}}^2
=a_l^2\phi_{l}^TP_{l-1}(P_{l-1}^{-1}+\beta_{l}\phi_{l}\phi_{l}^{T})
P_{l-1}\phi_{l}\tilde{s}_l^2\nonumber\\
&=a_l^2\phi_{l}^TP_{l-1}\phi_{l}(1+\beta_{l}\phi_{l}^{T}P_{l-1} \phi_{l})\tilde{s}_l^2
=a_l\phi_{l}^TP_{l-1}\phi_{l}\tilde{s}_l^2.\nonumber
\end{align}
Noting $P_l>0$, we have $\|P_l\|=\lambda_ {\max}(P_l)=\lambda_{\min}^{-1}(P_l^{-1})$.
By Assumptions \ref{AI} and \ref{AAB}, $0<a_l\leq 1$, $\|\hat{\theta}_{l}-\hat{\theta}_{l-1}\|^2\leq
{\|\hat{\theta}_{l}-\hat{\theta}_{l-1}\|_{P_{l}^{-1}}^2}/{\lambda_{\min}(P_l^{-1})} $, and Lemma \ref{prop:pk}, we can get
\vspace{-10pt}
\begin{align}
&\|\hat{\theta}_{l}-\hat{\theta}_{l-1}\|
\leq\sqrt{{a_l\phi_{l}^TP_{l-1}\phi_{l}\tilde{s}_l^2}/
{\lambda_{\min}(P_l^{-1})}}\nonumber\\
\leq & \sqrt{a_l\phi_{l}^TP_{l-1}\phi_{l}}|\tilde{s}_l| \|P_l\|^{\frac{1}{2}}
\leq2|\bar{\alpha}|\bar{\phi}\|P_{l-1}\|
\leq2|\bar{\alpha}|\bar{\phi}\|P_{k}\|.\nonumber
\end{align}
Then, taking it into (\ref{pe1}) yields this lemma.
\qed

{\bf Proof of Theorem \ref{thm:cov}:}

The proof is mainly based on Lyapunov function method, which is divided into the following three parts.

\textbf{Part I}: The mean square convergence properties.

Denote Lyapunov function as $V_k=\tilde\theta_k^TP_{k}^{-1}\tilde\theta_{k}$.
From (\ref{update}), (\ref{ak}) and Lemma \ref{lem:proj}, we have
\vspace{-10pt}
\begin{align}\label{st}
V_k=&\left(\Pi_{P_{k}^{-1}}\(\hat\theta_{k-1}+a_kP_{k-1}\phi_{k}\tilde{s}_k\)-\theta\right)^TP_{k}^{-1}\nonumber\\
&\cdot\left(\Pi_{P_{k}^{-1}}\(\hat\theta_{k-1}+a_kP_{k-1}\phi_{k}\tilde{s}_k\)-\theta\right)\nonumber\\
\leq & \(\tilde\theta_{k-1}+a_kP_{k-1}\phi_{k}\tilde{s}_k\)^T P_{k}^{-1}\(\tilde\theta_{k-1}+a_kP_{k-1}\phi_{k}\tilde{s}_k\)\nonumber\\
\leq & \tilde\theta_{k-1}^TP_{k-1}^{-1}\tilde\theta_{k-1}
+\beta_k\tilde\theta_{k-1}^T\phi_{k}\phi_{k}^T\tilde\theta_{k-1}
+2a_k\phi_{k}^T\tilde\theta_{k-1}\tilde{s}_k\nonumber\\
&+a_k^2\phi_{k}^TP_{k-1}\phi_{k}\tilde{s}_k^2
+2\beta_ka_k\phi_{k}^TP_{k-1}\phi_{k}\phi_{k}^T\tilde\theta_{k-1}\tilde{s}_k\nonumber\\
&+a_k^2\beta_k\phi_{k}^TP_{k-1}\phi_{k}\phi_{k}^TP_{k-1}\phi_{k}\tilde{s}_k^2\nonumber\\
\leq &V_{k-1}+\beta_k\tilde\theta_{k-1}^T\phi_{k}\phi_{k}^T \tilde\theta_{k-1}+2\phi_{k}^T\tilde\theta_{k-1}\tilde{s}_k\nonumber\\
&+a_k\phi_{k}^TP_{k-1}\phi_{k}\tilde{s}_k^2.
\end{align}
By $\mathbb{E }s_k
= \sum_{i=1}^{m} \alpha_{i,k} H_{i,k}$, $\hat{F}_{0,k}=F_{0,k}=0$, $\hat{F}_{m+1,k}=F_{m+1,k}=1$ and the differential mean value theorem,
\vspace{-10pt}
\begin{align}
&\mathbb{E}[\tilde s_k|\mathcal{F}_{k-1}]=\sum_{i=1}^{m+1} \alpha_{i,k} \(H_{i,k}- \hat{H}_{i,k}\)\nonumber\\
=&\sum_{i=1}^{m} (\alpha_{i+1,k}-\alpha_{i,k})\(F(C_i-\phi^T_{k}\hat\theta_{k-1}) -F\(C_i-\phi^T_{k}\theta\)\)\nonumber
\end{align}
\begin{align}\label{etildesk}
=&-\sum_{i=1}^{m} (\alpha_{i+1,k}-\alpha_{i,k}) f(C_i-\phi_k^T\xi_{i,k})\phi_k^T\tilde\theta_{k-1} \nonumber\\
=&-\sum_{i=1}^{m} (\alpha_{i+1,k}-\alpha_{i,k}) \check{f}_{i,k}\phi_k^T\tilde\theta_{k-1}
\end{align}
where $\xi_{i,k}$ with $\phi_k^T\xi_{i,k}$ in the interval between $\phi^T_{k}\theta$ and $\phi^T_{k}\hat\theta_{k-1}$ such that $F(C_i-\phi^T_{k}\hat\theta_{k-1})-
F(C_i-\phi^T_{k}\theta)=-f(C_i-\phi_k^T\xi_{i,k})
\phi_k^T\tilde\theta_{k-1}$,
and $\check{f}_{i,k}\triangleq f(C_i-\phi_k^T\xi_{i,k})$.
Then from (\ref{st})-(\ref{etildesk}) and $|\tilde{s}_k|\leq 2\bar{\alpha}$, we have
\vspace{-5pt}
\begin{align}\label{kk-h}
&\mathbb{E}V_k
\leq\mathbb{E}V_{k-1}+\mathbb{E}\beta_k\tilde\theta_{k-1}^T\phi_{k}\phi_{k}^T\tilde\theta_{k-1}
+\mathbb{E}a_k\phi_{k}^TP_{k-1}\phi_{k}\tilde{s}_k^2\nonumber\\
&\qquad\quad-2\sum_{i=1}^{m}\mathbb{E}(\alpha_{i+1,k}-\alpha_{i,k})\check{f}_{i,k}
\tilde\theta_{k-1}^T\phi_{k}\phi_{k}^T\tilde\theta_{k-1}\nonumber\\
&\leq\mathbb{E}V_{k-1}+\mathbb{E}\(1-\frac{2\sum_{i=1}^{m}(\alpha_{i+1,k}-\alpha_{i,k})
\check{f}_{i,k}}{\beta_k}\)\nonumber\\
&\quad\cdot\beta_k\tilde\theta_{k-1}^T\phi_{k}\phi_{k}^T\tilde\theta_{k-1}
+4\overline\alpha^2\mathbb{E}a_k\phi_{k}^TP_{k-1}\phi_{k}\nonumber\\
&\leq\mathbb{E}\tilde\theta_{k-1}^TP_{k-1}^{-\frac{1}{2}}\(I_n+
\(1-\frac{2\sum_{i=1}^{m}(\alpha_{i+1,k}-\alpha_{i,k})
\check{f}_{i,k}}{\beta_k}\)P_{k-1}^{\frac{1}{2}}\right. \nonumber\\
&\quad\cdot \left.\beta_k\phi_{k}\phi_{k}^T
P_{k-1}^{\frac{1}{2}}\)P_{k-1}^{-\frac{1}{2}}\tilde\theta_{k-1}
+4\overline\alpha^2 a_k\phi_{k}^TP_{k-1}\phi_{k}.
\end{align}
Denote
\vspace{-8pt}
\begin{align}\label{uf}
\underline f=\min_{1\leq i\leq m}
\min_{x\in[C_i-\bar{\phi}\bar{\theta},C_i+\bar{\phi}\bar{\theta}]}f(x).
\end{align}
From Assumption \ref{AN}, we have $\underline f>0$ and $\check{f}_{i,k}\geq\underline f$.  By Assumption \ref{AAB} and (\ref{kk-h}), we get
\vspace{-5pt}
\begin{align}\label{kk-hh}
\mathbb{E}V_k
\leq&\mathbb{E}\tilde\theta_{k-1}^TP_{k-1}^{-\frac{1}{2}}\(I_n+
\(1-\frac{2\alpha\underline f}{\overline{\beta}}\)
P_{k-1}^{\frac{1}{2}}\beta_k\phi_{k}\phi_{k}^T
P_{k-1}^{\frac{1}{2}}\)\nonumber\\
&\cdot P_{k-1}^{-\frac{1}{2}}\tilde\theta_{k-1}
+4\overline\alpha^2 a_k\phi_{k}^TP_{k-1}\phi_{k},
\end{align}
where $1-{2\alpha\underline f}/{\overline{\beta}}\leq {1}/{n}$ from Assumption \ref{AAB} and (\ref{uf}).

Next, we show the mean square convergence of WQNP algorithm in two cases, $1-\frac{2\alpha\underline f}{\overline{\beta}}\leq0$ and  $1-\frac{2\alpha\underline f}{\overline{\beta}}>0$.\\
\textbf{Case I-1}: $1-\frac{2\alpha\underline f}{\overline{\beta}}\leq0$.\\
Noticing $P_{k}^{-1}=P_{k-1}^{-1}\(I_n+\beta_{k}P_{k-1}\phi_{k} \phi_{k}^{T}\)$, we have
$\left|P_{k}^{-1}\right|
=\left|P_{k-1}^{-1}\right|\(1+\beta_{k}\phi_{k}^{T}P_{k-1}\phi_{k}\)$ and
$a_k\beta_{k}\phi_{k}^{T}P_{k-1}\phi_{k}={\(\left|P_{k}^{-1}\right|-\left|P_{k-1}^{-1}\right|\)}
/{\left|P_{k}^{-1}\right|}$.
Then,
\vspace{-5pt}
\begin{align}\label{appp}
\sum_{l=1}^{k}a_l\beta_{l}\phi_{l}^{T}P_{l-1}\phi_{l}
&=\sum_{l=1}^{k}\frac{\left|P_{l}^{-1}\right|-\left|P_{l-1}^{-1}\right|}
{\left|P_{l}^{-1}\right|}
\leq \sum_{l=1}^{k}\int_{\left|P_{l-1}^{-1}\right|}^{\left|P_{l}^{-1}\right|}
\frac{dx}{x}\nonumber\\
&\leq \log \left|P_{k}^{-1}\right|-\log\left|P_{0}^{-1}\right|.
\end{align}
From Lemma \ref{lemma4_ZZG}, we have
\vspace{-10pt}
\begin{align}\label{appp2}
\sum_{l=1}^{k}\(\beta_{l}\phi_{l}^{T}P_{l-1}\phi_{l}\)^2&<\infty.
\end{align}
And by (\ref{kk-hh}) and (\ref{appp}), we get
\vspace{-8pt}
\begin{align}
\mathbb{E}V_k
\leq&\mathbb{E}V_{k-1}+4\overline\alpha^2 a_k\phi_{k}^TP_{k-1}\phi_{k}\nonumber\\
\leq&\mathbb{E}V_{0}+\sum_{l=1}^{k}\frac{4\overline\alpha^2}{\underline{\beta}}\sum_{l=1}^{k} a_l\beta_k\phi_{l}^TP_{l-1}\phi_{l}
=O\(\log \left|P_{k}^{-1}\right|\).\nonumber
\end{align}
Then, combining Lemma \ref{prop:pk} gives
\begin{align}\label{dengyu}
\mathbb{E}\tilde{\theta}_{k}^T\tilde{\theta}_{k}
&\leq\mathbb{E}{V_k}/{\lambda_{\min}\(P_{k}^{-1}\)}
=O\({\log k}/{k}\).
\end{align}
\textbf{Case I-2}: $1-\frac{2\alpha\underline f}{\overline{\beta}}>0$. In this case, from (\ref{kk-hh}) we have
\vspace{-10pt}
\begin{align}\label{kk-hh1}
&\mathbb{E}V_k
\leq\mathbb{E}\tilde\theta_{k-1}^TP_{k-1}^{-\frac{1}{2}}\(1+
\(1-\frac{2\alpha\underline f}{\overline{\beta}}\)
\beta_k\phi_{k}^TP_{k-1}\phi_{k}\)\nonumber\\
&\qquad\quad\cdot P_{k-1}^{-\frac{1}{2}}\tilde\theta_{k-1}
+4\overline\alpha^2 a_k\phi_{k}^TP_{k-1}\phi_{k}\nonumber\\
\leq&\(1+\(1-\frac{2\alpha\underline f}{\overline{\beta}}\)
\beta_k\phi_{k}^TP_{k-1}\phi_{k}\) \mathbb{E}V_{k-1}
+4\overline\alpha^2 a_k\phi_{k}^TP_{k-1}\phi_{k}\nonumber\\
\leq&\prod_{l=1}^{k}\(1+\(1-\frac{2\alpha\underline f}{\overline{\beta}}\)
\beta_l\phi_{l}^TP_{l-1}\phi_{l}\) \mathbb{E}V_{0}
+4\overline\alpha^2\sum_{l=1}^{k}\prod_{i=l+1}^{k} \nonumber\\
&\(1+\(1-\frac{2\alpha\underline f} {\overline{\beta}}\)
\beta_i\phi_{i}^TP_{i-1}\phi_{i}\)a_l\phi_{l}^TP_{l-1}\phi_{l}.
\end{align}
First, we estimate the first item on the right side of (\ref{kk-hh1}) by (\ref{appp}), (\ref{appp2}) and Lemma \ref{prop:pk}. By $0<a_k\leq1$, we have
\vspace{-10pt}
\begin{align}\label{kk-hh11}
&\prod_{l=1}^{k}\(1+\(1-\frac{2\alpha\underline f}{\overline{\beta}}\)
\beta_l\phi_{l}^TP_{l-1}\phi_{l}\) \nonumber\\
=&e^{\sum_{l=1}^{k}\log\(1+\(1-{2\alpha\underline f}/{\overline{\beta}}\)\beta_l\phi_{l}^TP_{l-1}\phi_{l}\)}\nonumber\\
\sim&e^{\(1-{2\alpha\underline f} /{\overline{\beta}}\) \sum_{l=1}^{k}\beta_l\phi_{l}^TP_{l-1}\phi_{l}}\nonumber\\
=&e^{\(1-\frac{2\alpha\underline f}{\overline{\beta}}\)\sum_{l=1}^{k}
a_l\beta_l\phi_{l}^TP_{l-1}\phi_{l}}\cdot e^{\(1-\frac{2\alpha\underline f}{\overline{\beta}}\)\sum_{l=1}^{k}
a_l(\beta_l\phi_{l}^TP_{l-1}\phi_{l})^2}\nonumber\\
\leq&e^{\(1-{2\alpha\underline f}/{\overline{\beta}}\)\(\log \left|P_{k}^{-1}\right| -\log\left|P_{0}^{-1}\right|\)}\cdot M\nonumber\\
=&M\({\left|P_{k}^{-1}\right|}/{\left|P_{0}^{-1}\right|}\)^{ \(1-{2\alpha\underline f}/{\overline{\beta}}\)},
\end{align}
where $M$ ia a constant related to (\ref{appp2}).\\
Then, we estimate the second item on the right side of (\ref{kk-hh1}). Noticing (\ref{appp2}) and (\ref{kk-hh11}), we have
\vspace{-5pt}
\begin{align}\label{kk-hh12}
&4\overline\alpha^2\sum_{l=1}^{k} \prod_{i=l+1}^{k}\(1+\(1-\frac{2\alpha\underline f}{\overline{\beta}}\)
\beta_i\phi_{i}^TP_{i-1}\phi_{i}\)a_l\phi_{l}^TP_{l-1}\phi_{l}\nonumber\\
\leq& \frac{4\overline\alpha^2}{\underline{\beta}}\sum_{l=1}^{k} \prod_{i=l+1}^{k}\(1+\(1-\frac{2\alpha\underline f}{\overline{\beta}}\)
\beta_i\phi_{i}^TP_{i-1}\phi_{i}\)a_l\beta_l\phi_{l}^TP_{l-1}\phi_{l}\nonumber\\
\leq & \frac{4M\overline\alpha^2}{\underline{\beta}}
\left|P_{k}^{-1}\right|^{\(1-\frac{2\alpha\underline f}{\overline{\beta}}\)} \sum_{l=1}^{k}\frac{\left|P_{l}^{-1}\right|-\left|P_{l-1}^{-1}\right|} {\left|P_{l}^{-1}\right|^{2-{2\alpha\underline f}/{\overline{\beta}}}}\nonumber\\
\leq & \frac{4M\overline\alpha^2}{\underline{\beta}\(1-\frac{2\alpha\underline f}{\overline{\beta}}\)}\(\frac{\left|P_{k}^{-1}\right|}{\left|P_{0}^{-1}\right|}\)^{ 1-{2\alpha\underline f}/{\overline{\beta}}}.
\end{align}
Then, taking (\ref{kk-hh11}) and (\ref{kk-hh12}) into (\ref{kk-hh1}) gives
\vspace{-5pt}
\begin{align}
\mathbb{E}V_k=O\(\left|P_{k}^{-1}\right|^{ 1-{2\alpha\underline f}/{\overline{\beta}}}\).\nonumber
\end{align}
Hence, for $1-{2\alpha\underline f}/{\overline{\beta}}>0$, combining Lemma \ref{prop:pk} gives
\vspace{-5pt}
\begin{align}\label{dayu}
\mathbb{E}\tilde{\theta}_{k}^T\tilde{\theta}_{k}
&\leq\mathbb{E}\frac{V_k}{\lambda_{\min}\(P_{k}^{-1}\)}
=O\(k^{n\(1-{2\alpha\underline f}/{\overline{\beta}}\)-1}\),
\end{align}
where Assumption \ref{AAB} assures $n\(1-{2\alpha\underline f}/{\overline{\beta}}\)-1<0$.\\
Therefore, combining (\ref{dengyu}) and (\ref{dayu}) yields
\vspace{-5pt}
\begin{align}\label{mscr}
\mathbb{E}\tilde\theta_k^T\tilde\theta_{k}
=\left\{ \begin{array}{ll}
O\(\frac{\log k}{k}\), & \ \hbox{if }~~  \frac{2\alpha\underline f}{\overline{\beta}}\geq1,\\
O\(k^{n\(1-{2\alpha\underline f}/{\overline{\beta}}\)-1}\), & \ \hbox{if }~~  \frac{2\alpha\underline f}{\overline{\beta}}<1.
\end{array} \right.
\end{align}
\textbf{Part II}: This part focuses on the convergence property of the WQNP algorithm in the high rank square.\\
When $r=2$, from (\ref{st}), we have
\vspace{-10pt}
\begin{align}
V_k^2\leq&
\(V_{k-1}+(\phi_{k}^T\tilde\theta_{k-1})^2
+2\phi_{k}^T\tilde\theta_{k-1}\tilde{s}_k
+a_k\phi_{k}^TP_{k-1}\phi_{k}\tilde{s}_k^2\)^2\nonumber\\
\leq &V_{k-1}^2+(\beta_k\tilde\theta_{k-1}^T\phi_{k} \phi_{k}^T\tilde\theta_{k-1})^2
+2(\phi_{k}^T\tilde\theta_{k-1})^2\tilde{s}_k^2\nonumber\\
&+a_k^2(\phi_{k}^TP_{k-1}\phi_{k})^2\tilde{s}_k^4
+4\beta_k(\tilde\theta_{k-1}^T\phi_{k})^3\tilde{s}_k\nonumber\\
&+2V_{k-1}\(\beta_k\tilde\theta_{k-1}^T\phi_{k}\phi_{k}^T\tilde\theta_{k-1}
+2\phi_{k}^T\tilde\theta_{k-1}\tilde{s}_k\)\nonumber\\
&+2V_{k-1}a_k\phi_{k}^TP_{k-1}\phi_{k}\tilde{s}_k^2\nonumber\\
&+a_k\phi_{k}^TP_{k-1}\phi_{k}\tilde{s}_k^2\(\beta_k(\tilde\theta_{k-1}^T\phi_{k})^2
+2\phi_{k}^T\tilde\theta_{k-1}\tilde{s}_k\).\nonumber
\end{align}
Noticing $|\tilde{s}_k|\leq (m+1)|\bar{\a}|$, $\|\phi_k\|\leq\bar\phi$,  (\ref{uf}), (\ref{mscr}), Assumption \ref{AP} and Lemma \ref{prop:pk}, we have
\vspace{-5pt}
\begin{align}\label{esst}
\mathbb{E}V_k^2
\leq &\mathbb{E}V_{k-1}^2
+2\mathbb{E}V_{k-1}(\beta_k-2\sum_{i=1}^{m}(\a_{i+1,l}-\a_{i,l})\check{f}_{i,l})\nonumber\\
&\cdot \tilde\theta_{k-1}^T\phi_{k}\phi_{k}^T\tilde\theta_{k-1}
+O\({\mathbb{E}\|\tilde\theta_{k-1}\|^2}\)\nonumber\\
\leq &\mathbb{E}V_{k-1}^2+2\(1-{2\alpha\underline f}/{\overline{\beta}}\)\mathbb{E}V_{k-1} \beta_k\tilde\theta_{k-1}^T\phi_{k}\phi_{k}^T\tilde\theta_{k-1}\nonumber\\
&+O\({\mathbb{E}\|\tilde\theta_{k-1}\|^2}\).
\end{align}
Next, we consider this problem from the following two cases, i.e.,
 $1-{2\alpha\underline f}/{\overline{\beta}}\leq0$ and $1-{2\alpha\underline f}/{\overline{\beta}}>0.$\\
\textbf{Case II-1}:$1-{2\alpha\underline f}/ {\overline{\beta}}\leq0$. From (\ref{mscr}) and (\ref{esst}), we have
\vspace{-5pt}
\begin{align}
\mathbb{E}V_k^2
\leq&\mathbb{E}V_{k-1}^2+O\({\mathbb{E}\|\tilde\theta_{k-1}\|^2}\)
\leq\mathbb{E}V_{k-1}^2+O\(\frac{\log k}{k}\)\nonumber\\
=&\mathbb{E}V_{0}+O\(\sum_{l=1}^{k}\frac{\log l}{l}\)
=O\(\log^2 k\),\nonumber
\end{align}
which together with Lemma \ref{prop:pk} yields
\vspace{-5pt}
\begin{align}\label{ssdengyu}
\mathbb{E}\|\tilde{\theta}_{k}\|^4
&\leq\frac{\mathbb{E}V_k^2}{\lambda_{\min}^2\(P_{k}^{-1}\)}
=O\(\frac{\log^2 k}{k^2}\).
\end{align}
\textbf{Case II-2}:$1-{2\alpha\underline f}/{\overline{\beta}}>0$. By (\ref{dayu}) and (\ref{esst}), we have
\vspace{-5pt}
\begin{align}\label{esst1}
\mathbb{E}V_k^2
\leq&\(1+2\(1-{2\alpha\underline f}/{\overline{\beta}}\)
\beta_k\phi_{k}^TP_{k-1}\phi_{k}\) \mathbb{E}V_{k-1}^2\nonumber\\
&+O\(k^{n\(1-{2\alpha\underline f}/{\overline{\beta}}\)-1}\)\nonumber\\
\leq&\prod_{l=1}^{k}\(1+2\(1-\frac{2\alpha\underline f}{\overline{\beta}}\)\beta_l\phi_{l}^TP_{l-1}\phi_{l}\) \mathbb{E}V_{0}^2\nonumber\\
&+O\( \sum_{1=1}^{k}\prod_{i=l+1}^{k}\(1+2\(1-\frac{2\alpha \underline f}{\overline{\beta}}\)\beta_i\phi_{i}^TP_{i-1}\phi_{i}\) \right.\nonumber\\
&\cdot \left.l^{n\(1-{2\alpha\underline f}/{\overline{\beta}}\)-1}\)
= O\(k^{2n\(1-{2\alpha\underline f}/{\overline{\beta}}\)}\).
\end{align}
For $1-{2\alpha\underline f}/{\overline{\beta}}>0$,  combining Lemma \ref{prop:pk} and (\ref{esst1}) gives
\vspace{-5pt}
\begin{align}\label{ssdayu}
\mathbb{E}\|\tilde{\theta}_{k}\|^4
&\leq\frac{\mathbb{E}V_k}{\lambda_{\min}^2\(P_{k}^{-1}\)}
=O\(k^{2n\(1-{2\alpha\underline f}/{\overline{\beta}}\)-2}\).
\end{align}
Therefore, from (\ref{ssdengyu}) and (\ref{ssdayu}), we have
\vspace{-5pt}
\begin{align}
\mathbb{E}\|\tilde\theta_k\|^2
=\left\{ \begin{array}{ll}
O\(\frac{\log^2 k}{k^2}\), & \ \hbox{if }~~  \frac{2\alpha\underline f}{\overline{\beta}}\geq1,\\
O\(k^{2n\(1-{2\alpha\underline f}/{\overline{\beta}}\)-2}\), & \ \hbox{if }~~ \frac{2\alpha\underline f}{\overline{\beta}}<1.
\end{array} \right.\nonumber
\end{align}
Similar, for any $r\geq3$, we can get
\vspace{-5pt}
\begin{align}
\mathbb{E}\|\tilde\theta_k\|^{2r}
=&\left\{ \begin{array}{ll}
O\(\frac{(\log k)^r}{k^r}\), & \ \hbox{if }~~  \frac{2\alpha\underline f}{\overline{\beta}} =1,\\
O\(k^{rn\(1-{2\alpha\underline f}/{\overline{\beta}}\)-r}\), & \ \hbox{if }~~  \frac{2\alpha\underline f}{\overline{\beta}} <1,
\end{array} \right.\nonumber
\end{align}
where Assumption \ref{AAB} keeps $rn\(1-{2\alpha\underline f}/{\overline{\beta}}\)-r<0$.\\
In summary, there exists $\mu<\infty$ such that (\ref{mcr}) holds for any $r\geq1$, which implies (\ref{mc}).\\
\textbf{Part III}: The almost sure convergence of WQNP algorithm is consider in this part.
Denote $\bar{V}_k=\frac{V_k}{\lambda_{\min}(P_k^{-1})}$.
By (\ref{kk-hh}) and Lemma \ref{prop:pk}, we have
\vspace{-5pt}
\begin{align}
\mathbb{E}[\bar{V}_k|\mathcal{F}_{k-1}]
\leq&\bar{V}_{k-1}+{2\overline\alpha^2 a_k\phi_{k}^TP_{k-1}\phi_{k}}/{\lambda_{\min}(P_{k-1}^{-1})}\nonumber\\
\leq& \bar{V}_{k-1}+O\({1}/{k^2}\), \text{ for  } 1-{2\alpha\underline f}/{\overline{\beta}}\leq 0; \nonumber\\
\mathbb{E}[\bar{V}_k|\mathcal{F}_{k-1}]
\leq&\bar{V}_{k-1}+\(1-\frac{2\alpha\underline f}{\overline{\beta}}\)
\frac{\tilde\theta_{k-1}^T\beta_k\phi_{k}\phi_{k}^T\tilde\theta_{k-1}} {\lambda_{\min}(P_{k}^{-1})}\nonumber\\
&+{2\overline\alpha^2 a_k\phi_{k}^TP_{k-1}\phi_{k}}/ {\lambda_{\min}(P_{k-1}^{-1})}\nonumber\\
\leq& \bar{V}_{k-1}+\(1-{2\alpha\underline f}/{\overline{\beta}}\){\bar{\beta}\bar{\phi}^2
\|\tilde\theta_{k-1}\|^2}/{k}\nonumber\\
&+O\({1}/{k^2}\), \text{ for } 1-{2\alpha\underline f}/{\overline{\beta}}>0.\nonumber
\end{align}
From (\ref{mscr}), we have $\mathbb{E}\frac{\|\tilde\theta_ {k-1}\|^2}{k} =O\(k^{-2+n\(1-{2\alpha\underline f}/{\overline{\beta}}\)}\)$ when $1-{2\alpha\underline f}/{\overline{\beta}}>0$.
From $\sum_{k=1}^{\infty}k^{-2+n\(1-{2\alpha\underline f}/{\overline{\beta}}\)}<\infty$, $\sum_{k=1}^{\infty}{1}/{k^2}<\infty$ and Lemma \ref{Lemma 1.2.2_Chen2002}, $\bar{V}_k$ converges almost surely to a bounded limit. From (\ref{dengyu}) and (\ref{dayu}), we have $\mathbb{E}\bar{V}_k\to 0,k\rightarrow\infty$. Then, there is a subsequence
of $\bar{V}_k$ that converges almost surely to 0.  Noticing $\|\tilde\theta_k\|^2\leq \bar{V}_k$, $\tilde\theta_k$ almost surely converges to 0.
\qed

\subsection{Proof of Theorem \ref{thm:covrok}}\label{proof:thm:covrok}
Since $\check{f}_{i,k}\geq \inf_k\min_{1\leq i\leq m}\min_{\vartheta\in\Omega}f(C_i-\phi_k^T\vartheta)\triangleq \underline{f_{\phi}}$,  noticing (\ref{kk-h}) we have
\vspace{-12pt}
\begin{align}\label{kk-hh3}
\mathbb{E}V_k
\leq&\mathbb{E}\tilde\theta_{k-1}^TP_{k-1}^{-\frac{1}{2}}\(I_n+
\(1-{2\alpha\underline{f_{\phi}}}/{\overline{\beta}}\)
P_{k-1}^{\frac{1}{2}}\beta_k\phi_{k}\phi_{k}^T
P_{k-1}^{\frac{1}{2}}\)\nonumber\\
&\cdot P_{k-1}^{-\frac{1}{2}}\tilde\theta_{k-1}
+4\overline\alpha^2 a_k\phi_{k}^TP_{k-1}\phi_{k}\nonumber\\
\leq&\mathbb{E}V_{k-h}-\({2\alpha\underline{f_{\phi}}}/
{\overline{\beta}} -1\)\sum_{l=k-h}^{k-1}\mathbb{E}
\tilde\theta_{l}^T\beta_{l+1}\phi_{l+1}
\phi_{l+1}^T\tilde\theta_{l}\nonumber\\
&+\sum_{l=k-h}^{k-1}4\overline\alpha^2 a_{l+1}\phi_{l+1}^TP_{l}\phi_{l+1},
\end{align}
where ${2\alpha\underline{f_{\phi}}}/{\overline{\beta}} -1>0$ by (\ref{abc}).
From Assumptions \ref{AP} and \ref{AI}, we have $\|\tilde\theta_ {l}\|\leq2\bar{\theta}$ and $\|\phi_l\|\leq\bar{\phi}$. For $l=k-h,\ldots,k-1$, using Lemmas \ref{prop:pk} and \ref{projerror} gives
\vspace{-10pt}
\begin{align}\label{kk-hh31}
&-\tilde\theta_{l}^T\beta_{l+1}\phi_{l+1}
\phi_{l+1}^T\tilde\theta_{l}\nonumber\\
=&-\tilde\theta_{k-h}^T\beta_{l+1}\phi_{l+1}
\phi_{l+1}^T\tilde\theta_{k-h}+2\tilde\theta_{l}^T\beta_{l+1}\phi_{l+1}
\phi_{l+1}^T(\tilde\theta_{l}-\tilde\theta_{k-h})\nonumber\\
&-(\tilde\theta_{l}-\tilde\theta_{k-h})^T\beta_{l+1}\phi_{l+1}
\phi_{l+1}^T(\tilde\theta_{l-1}-\tilde\theta_{k-h})\nonumber\\
\leq&-\tilde\theta_{k-h}^T\beta_{l+1}\phi_{l+1}
\phi_{l+1}^T\tilde\theta_{k-h}+2\tilde\theta_{l}^T\beta_{l+1}\phi_{l+1}
\phi_{l+1}^T(\tilde\theta_{l}-\tilde\theta_{k-h})\nonumber\\
=&-\tilde\theta_{k-h}^T\beta_{l+1}\phi_{l+1}
\phi_{l+1}^T\tilde\theta_{k-h}+O\({1}/{(k-h)}\).
\end{align}
By Assumption \ref{AAB}, we have
\vspace{-10pt}
\begin{align}\label{kk-hh32}
&\sum_{l=k-h}^{k-1}\beta_{l+1}\phi_{l+1}\phi_{l+1}^T\geq h\underline\beta\delta^2I_n
\geq \frac{h\underline\beta\delta^2}{\(\|P_0^{-1}\|+ \overline{\beta}\bar{\phi}^2\)k}\nonumber\\
&\cdot\(P_{0}^{-1}+\sum_{l=1}^{k}\beta_{l}\phi_{l}\phi_{l}^{T}\)
\geq \frac{h\underline\beta\delta^2P_{k-h}^{-1}} {\(\|P_0^{-1}\|+ \overline{\beta}\bar{\phi}^2\)k}.
\end{align}
Denote ${\gamma}=\frac{\underline\beta\delta^2}{\(\|P_0^{-1}\|+ \overline{\beta}\bar{\phi}^2\)}\(\frac{2\alpha\underline{f_{\phi}}} {\overline{\beta}} -1\)>0$. By Lemmas \ref{Lemma1_ZWZ2021} and \ref{prop:pk}, substituting (\ref{kk-hh31}) and (\ref{kk-hh32}) into (\ref{kk-hh3}) gives
\vspace{-5pt}
\begin{align}
\mathbb{E}V_k
\leq&\mathbb{E}V_{k-h}-\({2\alpha\underline{f_{\phi}}}
/{\overline{\beta}} -1\){h\underline\beta\delta^2}/ {\(\|P_0^{-1}\|+\overline{\beta}\bar{\phi}^2\)k}\nonumber\\
&\cdot\mathbb{E}\tilde\theta_{k-h}^TP_{k-h}^{-1} \tilde\theta_{k-h}+O\({1}/{(k-h)}\)\nonumber\\
=&\(1-{h\gamma}/{k}\)\mathbb{E}V_{k-h}+O\({1}/{(k-h)}\)\nonumber\\
=&\prod_{l=1}^{\left\lfloor\frac{k}{h}\right\rfloor-1}\(1-\frac{h\gamma}{k-lh}\)
\mathbb{E}V_{k-\left\lfloor\frac{k}{h}\right\rfloor h}\nonumber\\
&+O\(\sum_{l=1}^{\left\lfloor\frac{k}{h}\right\rfloor-1}
\prod_{q=0}^{l-1}\(1-\frac{h\gamma}{k-qh}\)\frac{1}{k-lh}\)\nonumber\\
=&O\({1}/{k^{\gamma}}\)+O\(1\)=O\(1\).\nonumber
\end{align}
Then, by Lemma \ref{prop:pk}, we have
\vspace{-10pt}
\begin{align}
\mathbb{E}\tilde{\theta}_{k}^T\tilde{\theta}_{k}
&\leq{\mathbb{E}V_k}/{\lambda_{\min}\(P_{k}^{-1}\)}
=O\({1}/{k}\).\nonumber
\end{align}
Thus, the WQNP algorithm has a mean square convergence rate as $O\(\frac{1}{k}\)$.

\subsection{Proof of Proposition \ref{thm:CRLB}}\label{proof:thm:CRLB}
Since the noises $\{d_k\}$ are i.i.d., we have
\vspace{-10pt}
\begin{align}
&\mathbb{P}\(s_1,s_2,\cdots,s_k|\theta\)=\prod_{l=1}^k\mathbb{P}\(s_l|\theta\)
=\prod_{l=1}^k\sum_{i=1}^{m+1}H_{i,l}I_{\{s_l=\alpha_{i,l}\}}.\nonumber
\end{align}
Denote the log-likelihood function as
\vspace{-10pt}
\begin{align}
l_{k}(\theta)&=\log \mathbb{P}\(s_1,s_2,\cdots,s_k|\theta\)
=\sum_{l=1}^k\log\mathbb{P}\(s_i|\theta\)\nonumber\\
&=\sum_{l=1}^k\sum_{i=1}^{m+1}\log(H_{i,l})I_{\{s_l=\alpha_{i,l}\}}.\nonumber
\end{align}
Noticing that
$\frac{\partial \log H_{i,l}}{\partial \theta}=-\frac{h_{i,l}}{ H_{i,l}}\phi_l$,
and continuing the partial process, we have
$\frac{\partial^2 \log H_{i,l}}{\partial \theta^2}
=\frac{h'_{i,l}H_{i,l}-h_{i,l}^2}{H_{i,l}^2}\phi_l\phi_l^T$,
where
$h'_{i,l}=f'(C_i-\phi_l^T\theta)-f'(C_{i-1}-\phi_l^T \theta)$ for $j=2,\ldots, m$ and
$h'_{1, i}= f'(C_1-\phi_l^T \theta)$, $h'_{m+1, i}= -
f'(C_m-\phi_l^T \theta)$ with $f'(x)=\partial f(x) /\partial x$.
 Hence, $\sum_{i=1}^{m+1} h'_{i,l}=0$ and
 \vspace{-10pt}
\begin{align}
&\frac{\partial^2 l_k}{\partial \theta \partial \theta}
= \sum_{l=1}^k\left[\sum_{i=1}^{m+1}\frac{h'_{i,l} H_{i,l}-h_{i,l}^2}{H^2_{i,
l}}I_{\{s_l=\alpha_{i,l}\}}\right]\phi_l\phi_l^T,\nonumber
\end{align}
together with $\mathbb{E}I_{\{s_l=\alpha_{i,l}\}}=H_{i,l}$, the CR lower bound is
\vspace{-10pt}
\begin{align}
\Delta_{k}=&\(-\mathbb{E}\frac{\partial^2 l_k}{\partial \theta^2 }\)^{-1}\nonumber\\
=& \(-\sum_{l=1}^k \left(\sum_{i=1}^{m+1}\frac{h'_{i,l}
H_{i,l}-h_{i,l}^2}{H^2_{i, l}}
\mathbb{E}I_{\{s_l=\alpha_{i,l}\}}\right)\phi_l\phi_l^T\)^{-1}\nonumber\\
=& \(-\sum_{l=1}^k \left(\sum_{i=1}^{m+1}\frac{h'_{i,l}
H_{i,l}-h_{i,l}^2 }{ H_{i, l}}\right)\phi_l\phi_l^T\)^{-1}\nonumber\\
=& \(\(-\sum_{l=1}^k \sum_{i=1}^{m+1} h'_{i,l} + \sum_{l=1}^k
\sum_{i=1}^{m+1} \frac{h_{i,l}^2 }{H_{i, l}}\)\phi_l\phi_l^T\)^{-1}\nonumber\\
=& \(\sum_{l=1}^k \sum_{i=1}^{m+1}\frac{h_{i,l}^2 }{H_{i,l}}
\phi_l\phi_l^T\)^{-1}.\nonumber
\end{align}

\subsection{Proof of Proposition \ref{thm:lambdaq}}\label{proof:thm:lambdaq}
Since $f'(x)=-\frac{x}{\sigma^2} f(x)$ for the normally density function $f(x)$ with covariance $\sigma^2$, we have\\
\vspace{-12pt}
\begin{align}
\lim_{\Delta C_i \to 0}\rho_l
 =&\lim_{\Delta C_i \to0}\sum_{i=1}^{m+1}\frac{ h_{i,l}^2}{H_{i,l}}
 =\int_{-\infty}^{\infty} \frac{\(f'(x)\)^2}{ f(x)} {\rm d} x \nonumber\\
 =&\int_{-\infty}^{\infty} \({-\frac{x }{\sigma^2}}\)^2  f(x) {\rm d} x
 = \frac{1}{\sigma^2}.\nonumber
\end{align}
Hence, Proposition \ref{thm:lambdaq} holds.

\subsection{Proof of Proposition \ref{thm:condlamb}}
\label{proof:thm:condlamb}
Before proving Propostion \ref{thm:condlamb}, we give the following lemma to analyzing the properties of $h_{i,k}$ and $H_{i,k}$.
\begin{lemma}\label{lem:dgx}
Let $g(x,y)=\left\{
   \begin{array}{ll}
     \frac{f(x)-f(y)}{F(x)-F(y)}, & \hbox{if } ~ x\neq y; \\
     -\frac{y}{\sigma^2}, & \hbox{if } x=y.
   \end{array}
 \right.$ Then, $g_x(x,y)<0$ when $x\neq y$.
\end{lemma}

{\bf Proof:}
 Denote $\bar{g}(x,y)= \frac{x }{\sigma^2}(F(x)-F(y)) + (f(x)-f(y)).$
Noticing that $\bar{g}(y,y)=0$ and
$\bar{g}_x'(x,y)={F(x)-F(y)}/{\sigma^2},$
we have $\bar{g}(x,y)>0$ when $x\neq y$. Since
$f'(x)=-{x}f(x)/{\sigma^2} ,$
we get
\vspace{-10pt}
\begin{align}
g_x'(x,y)&=\({(f(x)-f(y))}/{ (F(x)-F(y))}\)'_x\nonumber\\
&=\frac{-\frac{x}{\sigma^2}f(x)(F(x)-F(y))-f(x)(f(x)-f(y)}{ (F(x)-F(y))^2}\nonumber\\
&={-f(x)\bar{g}(x,y)}/{(F(x)-F(y))^2}.\nonumber
\end{align}
So, we have $g_x'(x,y)<0$ when $x\neq y$.
\qed

Based on Lemma \ref{lem:dgx}, we give the following lemma, which can lead to Proposition \ref{thm:condlamb} directly.

\begin{lemma}\label{lem:dechoH}
For $x\in (-\infty, \infty)$ and $i=1, \ldots, m+1$, denote
$h_i(x)=f(C_i -x)-f(C_{i-1}-x)$ and $H_i(x)=F(C_i -x)-F(C_{i-1}-x)$. Then, for $i=2, \ldots, m+1$,
\vspace{-8pt}
\begin{gather}\label{hoHix}
\frac{h_i(x)}{H_i(x)} <\frac{h_{i-1}(x)}{H_{i-1}(x)}.
\end{gather}
\end{lemma}
{\bf Proof:}
From Lemma \ref{lem:dgx} and $C_i > C_{i-1}> C_{i-2}$ for
$i=2, \ldots, m+1$, we have$\frac{f(C_i-x)-f(C_{i-1}-x)}{ F(C_i-x)-F(C_{i-1}-x)} < \frac{f(C_{i-2}-x)-f(C_{i-1}-x)}
{F(C_{i-2}-x)-F(C_{i-1}-x)}$,
 which is equivalent to (\ref{hoHix}).
\qed

\subsection{Proof of Theorem \ref{thm:opt-cov}}\label{proof:thm:opt-cov}
From the definition of $H_{i,k}$ and $\hat{H}_{i,k}$ in (\ref{realHh}) and (\ref{estHh}), there exists  $\grave{\theta}_{i,k-1}$ with $\phi^T_{k}\grave{\theta}_{i,k-1}$ in the interval between $\phi^T_{k}\theta$ and $\phi^T_{k}\hat\theta_{i,k-1}$ such that
\vspace{-10pt}
\begin{align}\label{eetildesk}
\mathbb{E}[\tilde s_k|\mathcal{F}_{k-1}]&=\sum_{i=1}^{m+1} \hat\alpha_{i,k} \(H_{i,k}- \hat{H}_{i,k}\)\nonumber\\
&=\sum_{i=1}^{m} \(\hat\alpha_{i+1,k}-\hat\alpha_{i,k}\)\(\hat F_{i,k}- F_{i,k}\)\nonumber\\
&=-\sum_{i=1}^{m} \(\hat\alpha_{i+1,k}-\hat\alpha_{i,k}\)
\grave{f}_{i,k}\phi_k^T\tilde\theta_{k-1},
\end{align}
where $\grave{f}_{i,k}\triangleq f(C_i-\phi^T_ {k}\grave{\theta}_{i,k-1})\geq\min_{x\in[C_i-\bar\phi\bar\theta ,C_i+\bar\phi\bar\theta]}f(x)$.
Denote
\vspace{-10pt}
\begin{align}\label{lambda_k}
\lambda_k=\sum_{i=1}^{m} \(\hat\alpha_{i+1,k}-\hat\alpha_{i,k}\)
\grave{f}_{i,k}/{\hat{\beta}_k}.
\end{align}
By the continuity of $f(x)$ and $F(x)$, $\lambda_k$ and $\hat\beta_k$ are bounded.
From (\ref{estHh}), (\ref{opt:ab}), (\ref{ANN}) and (\ref{lambda_k}),
\vspace{-10pt}
\begin{align}\label{lambda_k_u}
\lambda_k&=\frac{\sum_{i=1}^{m} \(\hat\alpha_{i+1,k}-\hat\alpha_{i,k}\)
\grave{f}_{i,k}}{\sum_{i=1}^{m+1}\frac{\hat{h}^2_{i,k}}{\hat{H}_{i,k}}}
=\frac{\sum_{i=1}^{m} \(\hat\alpha_{i+1,k}-\hat\alpha_{i,k}\)
\grave{f}_{i,k}}{\sum_{i=1}^{m} \(\hat\alpha_{i+1,k}-\hat\alpha_{i,k}\)
\hat{f}_{i,k}}\nonumber\\
&\geq \frac{\sum_{i=1}^{m} \(\hat\alpha_{i+1,k}-\hat\alpha_{i,k}\)
\min_{x\in[C_i-\bar\phi\bar\theta,C_i+\bar\phi\bar\theta]}f(x)} {\sum_{i=1}^{m} \(\hat\alpha_{i+1,k}-\hat\alpha_{i,k}\)
\max_{x\in[C_i-\bar\phi\bar\theta,C_i+\bar\phi\bar\theta]}f(x)}
\geq \frac{1}{2}.
\end{align}
Let
\vspace{-8pt}
\begin{align}
&\underline{\lambda}=\inf_k\lambda_k, \overline{\lambda}=\sup_k\lambda_k, \overline{\hat\alpha}=\sup_k\max_{i=1,\ldots,m+1}
|\hat\alpha_{i,k}|;\label{b_lambda}\\
 &\underline{\hat\beta}=\inf_k\hat\beta_k, \overline{\hat\beta}=\sup_k\hat\beta_k.\label{b_beta}
\end{align}
Then, it can be seen that $\underline{\lambda}>1/2$, $\underline{\hat\beta}>0$, $\overline{\lambda}<\infty$ and  $\overline{\hat\beta}<\infty$ from the boundness of $\hat\theta_{k}$ and $\phi_{k}$.\\
Let $\hat{V}_k=\tilde\theta_k^T\hat{P}_{k}^{-1}\tilde\theta_{k}$. Similar to (\ref{st}), we have
\vspace{-10pt}
\begin{align}\label{opt:st}
\hat{V}_k\leq &\hat V_{k-1}+\hat \beta_k(\phi_{k}^T\tilde\theta_{k-1})^2
+2\phi_{k}^T\tilde\theta_{k-1}\tilde{s}_k\nonumber\\
&+\hat a_k\phi_{k}^T\hat P_{k-1}\phi_{k}\tilde{s}_k^2.
\end{align}
By (\ref{eetildesk})-(\ref{opt:st}), we have
\vspace{-8pt}
\begin{align}\label{opt-kk-h}
&\mathbb{E}\hat{V}_k\leq \mathbb{E}\hat V_{k-1}+\mathbb{E}\hat \beta_k\tilde\theta_{k-1}^T\phi_{k}\phi_{k}^T\tilde\theta_{k-1}
+\mathbb{E}\hat a_k\phi_{k}^T\hat P_{k-1}\phi_{k}\tilde{s}_k^2\nonumber\\
&\qquad\quad+2\mathbb{E}\sum_{i=1}^{m+1} \hat\alpha_{i,k} \(H_{i,k}- \hat{H}_{i,k}\)\phi_{k}^T\tilde\theta_{k-1}\nonumber\\
\leq&\mathbb{E}\hat V_{k-1}+\mathbb{E}\(1-2\lambda_k\)\hat \beta_k\tilde\theta_{k-1}^T\phi_{k}\phi_{k}^T\tilde\theta_{k-1}
+\mathbb{E}\hat a_k\phi_{k}^T\hat P_{k-1}\phi_{k}\tilde{s}_k^2\nonumber\\
\leq&\mathbb{E}\hat V_{k-1}+\mathbb{E}\(1-2\underline{\lambda}\)\hat \beta_k\tilde\theta_{k-1}^T\phi_{k}\phi_{k}^T\tilde\theta_{k-1}\nonumber\\
&+\mathbb{E}\hat a_k\phi_{k}^T\hat P_{k-1}\phi_{k}\tilde{s}_k^2,
\end{align}
where $1-2\underline{\lambda}\leq0$.\\
{Next, we discuss the convergence rate based on the higher moments and cross product of  estimation errors.\\
First, we show the mean square convergence rate of IBID algorithm can reach $O\(\frac{\log k}{k}\)$.} Similar to (\ref{appp}), we have
\vspace{-14pt}
\begin{align}
&\sum_{l=1}^{k}\hat a_l\hat \beta_{l}\phi_{l}^{T}\hat P_{l-1}\phi_{l}\leq \log \left|\hat P_{k}^{-1}\right|-\log\left|\hat P_{0}^{-1}\right|.\label{opt:appp1}
\end{align}
Noticing $|\tilde{s}_k|\leq 2\overline{\hat\alpha}$, (\ref{opt-kk-h}) and (\ref{opt:appp1}), we have
\vspace{-8pt}
\begin{align}
\mathbb{E}\hat V_k
\leq&\mathbb{E}\hat V_{k-1}+\mathbb{E}\hat a_k\phi_{k}^T\hat P_{k-1}\phi_{k}\tilde{s}_k^2\nonumber\\
\leq&\mathbb{E}\hat V_{0}+ \frac{4\overline{\hat\alpha}^2}{\underline{\hat\beta}}\sum_{l=1}^{k} \mathbb{E}\hat a_l\hat\beta_k\phi_{l}^T\hat P_{l-1}\phi_{l}
=O\(\log \mathbb{E}\left|\hat P_{k}^{-1}\right|\).\nonumber
\end{align}
From (\ref{b_beta}) and Assumption \ref{AI}, we get
\vspace{-8pt}
\begin{align}\label{hatP}
\hat{P}_k=O\({1}/{k}\) \text{ and } \hat{P}_k^{-1}=O\(k\).
\end{align}
From (\ref{hatP}), we have
\vspace{-8pt}
\begin{align}\label{opt:dengyu}
\mathbb{E}\tilde{\theta}_{k}^T\tilde{\theta}_{k}
&\leq\mathbb{E}{\hat V_k}/{\lambda_{\min}\(\hat P_{k}^{-1}\)}
=O\({\log k}/{k}\).
\end{align}
{Second, we establish the higher moments convergence rate of estimation errors (i.e., $\mathbb{E}\|\tilde\theta_k\|^{2r},r\geq2$) similarly to Part II in the proof of Theorem \ref{thm:cov}.}\\
Based on (\ref{opt:st}), (\ref{opt-kk-h}) and (\ref{opt:dengyu}), similar to (\ref{esst}) we can get
\vspace{-10pt}
\begin{align}
&\mathbb{E}\hat V_k^2
\leq \mathbb{E}\hat V_{k-1}^2
+2\mathbb{E}\hat V_{k-1}(\beta_k-2\sum_{i=1}^{m} (\hat\a_{i+1,l}-\hat\a_{i,l}) \grave{f}_{i,l})\nonumber\\
&\qquad\quad\cdot \tilde\theta_{k-1}^T\phi_{k}\phi_{k}^T \tilde\theta_{k-1}
+O\({\mathbb{E}\|\tilde\theta_{k-1}\|^2}\)\nonumber\\
\leq &\mathbb{E}\hat V_{k-1}^2+2\(1-2\underline{\lambda}\) \mathbb{E}\hat V_{k-1} \beta_k\tilde\theta_{k-1}^T\phi_{k}\phi_{k}^T \tilde\theta_{k-1}+O\({\mathbb{E}\|\tilde\theta_{k-1}\|^2}\)\nonumber\\
\leq &\mathbb{E}\hat V_{k-1}^2+O\(\frac{\log k}{k}\)
\leq \mathbb{E}\hat V_{0}^2+O\(\sum_{l=1}^{k}\frac{\log l}{l}\)
=O\({\log^2 k}\),\nonumber
\end{align}
which together with Lemma \ref{prop:pk} yields
\vspace{-8pt}
\begin{align}
\mathbb{E}\|\tilde{\theta}_{k}\|^4
&\leq{\mathbb{E}\hat V_k^2}/{\lambda_{\min}^2\(P_{k}^{-1}\)}
=O\({\log^2 k}/{k^2}\).\nonumber
\end{align}
Similar, for any $r\geq 1$, we can get
\vspace{-8pt}
\begin{align}\label{opt:esstm}
\mathbb{E}\|\tilde\theta_k\|^{2r}=
O\({(\log k)^r}/{k^r}\), \forall r=1,2,3\ldots
\end{align}
{Third, we will give the recursive form of the cross product of the estimation errors, i.e., $\mathbb{E}\tilde\theta_k\tilde\theta_k^T$, and then, we will prove that the mean square convergence rate of the IBID algorithm reaches $O\(\frac{1}{k}\)$.
Let $\theta_k=\hat\theta_ {k-1}+\hat{a}_k\hat{P}_{k-1}\phi_{k}\tilde{s}_k$ and $\bar{\theta}_k=\theta_k-\theta$. Then, $\hat\theta_k=\Pi_ {P_k^{-1}}\(\theta_k\)$ and
\vspace{-10pt}
\begin{align}\label{bart}
\bar{\theta}_k=\tilde\theta_{k-1}+\hat a_k\hat{P}_{k-1}\phi_{k}\tilde{s}_k.
\end{align}
Based on (\ref{hatP}), (\ref{opt:esstm}) and (\ref{bart}), we have
\vspace{-8pt}
\begin{align}\label{wpcr}
\mathbb{E}\|\bar\theta_k\|^{2r}=O\({(\log k)^r}/{k^r}\),r=1,2,3\ldots
\end{align}
Without loss of generality, we assume $\theta\in\Omega-\partial\Omega$, where $\partial\Omega$ is  the edge set of $\Omega$. Denote $\underline{\omega}=\min_{\omega\in\partial\Omega}\|\omega-\theta\|>0$. Then by Markov inequality,
\vspace{-8pt}
\begin{align}\label{wpp}
\mathbb{P}\(\theta_k\notin\Omega\)
&\leq\mathbb{P}\(\|\theta_k-\theta\|\geq\underline{\omega}\)
=\mathbb{P}\(\|\bar\theta_k\|\geq\underline{\omega}\)\nonumber\\
&=\mathbb{P}\(\|\bar\theta_k\|^{2r}\geq\underline{\omega}^{2r}\)
\leq{\mathbb{E}\|\bar\theta_k\|^{2r}}/{\underline{\omega}^{2r}}.
\end{align}
Noticing  $\|\bar{\theta} _k-\tilde{\theta}_k\|=0$ when ${\theta}_k\in\Omega$, and $\|\bar{\theta}_k-\tilde{\theta}_k\|\leq
\|\hat{a}_k\hat{P}_{k-1}\phi_{k}\tilde{s}_k\|=O\(\frac{1}{k}\)$ when ${\theta}_k\notin\Omega$, we have
\vspace{-8pt}
\begin{align}\label{wj1}
\mathbb{E}(\bar\theta_k-\tilde\theta_k)(\bar\theta_k-\tilde\theta_k)^T
&\leq \mathbb{E}\|\bar\theta_k-\tilde\theta_k\|^2I_n\nonumber\\
&\leq O\(1/k^{2}\)\cdot \mathbb{P}\(\theta_k\notin\Omega\).
\end{align}
For $a\in\mathbb{R}^n$ and $b\in\mathbb{R}^n$, we have $ab^T+ba^T\leq 2\sqrt{a^Tab^Tb}I_n$. Then, from (\ref{wpcr}), (\ref{wpp}) and (\ref{wj1}), we have
\vspace{-5pt}
\begin{align}\label{wj2}
&\mathbb{E}\tilde\theta_k\tilde\theta_k^T
=\mathbb{E}\bar\theta_k\bar\theta_k^T
+\mathbb{E}(\tilde\theta_k-\bar\theta_k)\bar\theta_k^T
+\mathbb{E}\bar\theta_k(\tilde\theta_k-\bar\theta_k)^T\nonumber\\
&\qquad\quad+\mathbb{E}(\bar\theta_k-\tilde\theta_k)(\bar\theta_k-\tilde\theta_k)^T\nonumber\\
\leq& \mathbb{E}\bar\theta_k\bar\theta_k^T +2\sqrt{\mathbb{E}\bar\theta_k^T\bar\theta_k\cdot\mathbb{E} (\bar\theta_k-\tilde\theta_k)^T(\bar\theta_k-\tilde\theta_k)}I_n\nonumber\\
\leq& \mathbb{E}\bar\theta_k\bar\theta_k^T
+O\({1}/{k^{3/2}}\)\sqrt{\mathbb{P}\(\theta_k\notin\Omega\)}
+ O\(1/k^{2}\)\cdot \mathbb{P}\(\theta_k\notin\Omega\)\nonumber\\
 =&\mathbb{E}\bar\theta_k\bar\theta_k^T+o\({1}/{k^2}\).
\end{align}
By (\ref{estHh}) and (\ref{opt:alpha}), we have $\mathbb{E}\left[\tilde{s}_k^2\big|\mathcal{F} _{k-1}\right] =\sum_{i=1}^{m+1}\hat\alpha_{i,k}^2{H}_{i,k}$ and
$\sum_{i=1}^{m+1} \hat\alpha_{i,k}\hat{H}_{i,k}=0$.
Then, by (\ref{eetildesk}), (\ref{hatP}), (\ref{bart}), (\ref{wj2}), $\mathbb{E}[\tilde{s}_k|\mathcal{F}_{k-1}] =\sum_{i=1}^{m+1}\hat\alpha_{i,k}{H}_{i,k}$ and Assumptions \ref{AP}-\ref{AI},
\vspace{-10pt}
\begin{align}\label{wj3}
&\mathbb{E}\tilde\theta_k\tilde\theta_k^T
\leq \mathbb{E}\tilde\theta_{k-1}\tilde\theta_{k-1}^T+ \mathbb{E}\sum_{i=1}^{m+1}\hat\alpha_{i,k}^2{H}_{i,k} \hat{a}_k^2\hat{P}_{k-1}\phi_{k}\phi_{k}^T\hat{P}_{k-1}\nonumber\\
&\quad+\mathbb{E}\sum_{i=1}^{m+1}\hat\alpha_{i,k}({H}_{i,k}-\hat{H}_{i,k}) \hat{a}_k\tilde\theta_{k-1}\phi_{k}^T\hat{P}_{k-1}\nonumber\\
&\quad+\mathbb{E}\sum_{i=1}^{m+1}\hat\alpha_{i,k}({H}_{i,k}-\hat{H}_{i,k}) \hat{a}_k\hat{P}_{k-1}\phi_{k}\tilde\theta_{k-1}^T+o\(\frac{1}{k^{2}}\)\nonumber\\
&\leq \mathbb{E}\tilde\theta_{k-1}\tilde\theta_{k-1}^T
-\mathbb{E}\tilde\theta_{k-1}\tilde\theta_{k-1}^T{a}_k\beta_k\phi_{k} \phi_{k}^T{P}_{k-1}\nonumber\\
&\quad-\mathbb{E}{a}_k{P}_{k-1}\beta_k\phi_{k}\phi_{k}^T\tilde\theta_{k-1}\tilde\theta_{k-1}^T
+\mathbb{E}\tilde\theta_{k-1}\tilde\theta_{k-1}^T\phi_{k}\phi_{k}^T\nonumber\\
&\quad\quad\cdot\big(\beta_{k}{a}_kP_{k-1}
-\sum_{i=1}^{m} \(\hat\alpha_{i+1,k}-\hat\alpha_{i,k}\)
\grave{f}_{i,k}\hat{a}_k\hat{P}_{k-1}\big)\nonumber\\
&\quad+\mathbb{E}\big(\beta_{k}{a}_kP_{k-1}
-\sum_{i=1}^{m} \(\hat\alpha_{i+1,k}-\hat\alpha_{i,k}\)
\grave{f}_{i,k}\hat{a}_k\hat{P}_{k-1}\big)\nonumber\\
&\quad\quad\cdot\phi_{k}\phi_{k}^T\tilde\theta_{k-1}\tilde\theta_{k-1}^T +O\(1/{k^{2}}\),
\end{align}
where $P_k$ is generated by (\ref{variance}) with ${\beta} _k=\sum_{i=1}^{m+1}\frac{{h}^2_{i,k} }{{H}_{i,k}}$,
$\alpha_{i,k}=-\frac{{h}_{i,k}}{{H}_{i,k}}$ and $a_k=\(1+\beta_k\phi_ {k}^TP_{k-1}\phi_{k}\)^{-1}$.
Then, by ${f}_{m+1,k}={f}_{0,k}=0$ and (\ref{realHh}), we have ${\beta}_k=-\sum_{i=1}^{m+1}\alpha_{i,k}\({f}_{i,k}-{f}_{i-1,k}\)
=\sum_{i=1}^{m}\(\alpha_{i+1,k}-\alpha_{i,k}\){f}_{i,k}.$
From Assumptions \ref{AI}, we have
\vspace{-5pt}
 \begin{align}\label{Pk}
 {P}_k=O\({1}/{k}\)  \text{ and } {P}_k^{-1}=O\(k\).
 \end{align}
Denote $\a_i(x)=-\frac{f(C_i-x)-f(C_{i-1}-x)} {F(C_i-x)-F(C_{i-1}-x)}$.
Then, $\alpha_{i,k}=\a_i(\phi_k\theta),$ $\hat\alpha_{i,k}=\a_i(\phi_k\hat\theta_{k-1})$.
From the continuous differentiability of $f(\cdot)$ and $F(\cdot)$, we get $\a_i(\cdot)$ is the continuous differentiable.
From (\ref{etildesk}), (\ref{hatP}),  (\ref{Pk}) and  $a_k, \hat{a}_k\in(0,1)$,
\vspace{-10pt}
\begin{align}\label{wj31}
&\left\|\beta_{k}{a}_kP_{k-1}
-\sum_{i=1}^{m}\(\hat\alpha_{i+1,k}-\hat\alpha_{i,k}\)
\grave{f}_{i,k}\hat{a}_k\hat{P}_{k-1}\right\|\nonumber\\
=&O\(\frac{1}{k}\)\cdot\left|\sum_{i=1}^{m}\(\alpha_{i+1,k}-\alpha_{i,k}\) {f}_{i,k}-\sum_{i=1}^{m}\(\hat\alpha_{i+1,k}-\hat\alpha_{i,k}\)
\grave{f}_{i,k}\right|\nonumber\\
=&O\(\frac{1}{k}\)\cdot\left|\sum_{i=1}^{m}\(\alpha_{i+1,k}-\alpha_{i,k}\) \({f}_{i,k}-\grave{f}_{i,k}\)\right|\nonumber\\
&+O\(\frac{1}{k}\)\cdot\left|\sum_{i=1}^{m+1}\(\alpha_{i,k}-\hat\alpha_{i,k}\)
\(\grave{f}_{i,k}-\grave{f}_{i-1,k}\)\right|\nonumber\\
=&O\(\frac{1}{k}\)\cdot\bigg|\sum_{i=1}^{m+1}\(\alpha_{i+1,k}-\alpha_{i,k}\)
f'(\grave{\zeta}_{i,k})\phi_k^T(\breve\theta_{k-1}-\theta)\bigg|\nonumber\\
&+O\(\frac{1}{k}\)\cdot\bigg|\sum_{i=1}^{m+1} {\alpha}_i^{'}(\hat{\xi}_{i,k})\phi_k^T\tilde\theta_k \(\grave{f}_{i,k}-\grave{f}_{i-1,k}\)\bigg|\nonumber\\
=&O\({1}/{k}\)\cdot\|\tilde\theta_k\|,
\end{align}
where $\hat{\xi}_{i,k}$ is between $\phi^T_ {k}\theta$ and $\phi^T_{k}\hat\theta_{k-1}$, $\grave{\zeta}_{i,k}$ is between $C_i-\phi^T_ {k}\grave\theta_{i,k-1}$ and $C_i-\phi^T_{k}\theta$,
$\grave{f}_{i,k}$ and $\grave\theta_{i,k-1}$ are denoted as (\ref{eetildesk}). Then,  based on (\ref{wpcr}), we have
\vspace{-10pt}
\begin{align}\label{wj4}
&\mathbb{E}\(\beta_{k}{a}_kP_{k-1}
-\sum_{i=1}^{m} \(\hat\alpha_{i+1,k}-\hat\alpha_{i,k}\)
\grave{f}_{i,k}\hat{a}_k\hat{P}_{k-1}\)\nonumber\\
&\cdot\tilde\theta_{k-1}\tilde\theta_{k-1}^T\phi_{k}\phi_{k}^T+\mathbb{E}\phi_{k}\phi_{k}^T\tilde\theta_{k-1}
\tilde\theta_{k-1}^T \nonumber\\
&\cdot\(\beta_{k}{a}_kP_{k-1}
-\sum_{i=1}^{m} \(\hat\alpha_{i+1,k}-\hat\alpha_{i,k}\)
\grave{f}_{i,k}\hat{a}_k\hat{P}_{k-1}\)\nonumber\\
\leq&O\(\frac{1}{k}\)\cdot\mathbb{E}\|\tilde\theta_k\|^3
\leq O\(\frac{1}{k}\)\cdot\sqrt{\mathbb{E}\|\tilde\theta_k\|^2\cdot \mathbb{E}\|\tilde\theta_k\|^4}\nonumber\\
=&O\({1}/{k}\)\cdot\sqrt{O\({\log^3 k}/{k^3}\)}
=o\({1}/{k^2}\).
\end{align}
By (\ref{Pk}) and $P_k=P_{k-1}-a_kP_{k-1}\beta_k\phi_{k}\phi_{k}^TP_{k-1}$, taking (\ref{wj4}) into (\ref{wj3}) yields
\vspace{-10pt}
\begin{align}
\mathbb{E}\tilde\theta_k\tilde\theta_k^T
\leq& \mathbb{E}\tilde\theta_{k-1}\tilde\theta_{k-1}^T
-\mathbb{E}\tilde\theta_{k-1}\tilde\theta_{k-1}^T{a}_k\beta_k \phi_{k}\phi_{k}^T{P}_{k-1}\nonumber\\
&-\mathbb{E}{a}_k{P}_{k-1}\beta_k\phi_{k}\phi_{k}^T\tilde\theta_{k-1}\tilde\theta_{k-1}^T
+O\({1}/{k^{2}}\)\nonumber\\
\leq& (I_n-{a}_k{P}_{k-1}\beta_k\phi_{k}\phi_{k}^T) \mathbb{E}\tilde\theta_{k-1}\tilde\theta_{k-1}^T\nonumber\\
&\cdot(I_n-{a}_k\beta_k \phi_{k}\phi_{k}^T{P}_{k-1}) +O\({1}/{k^{2}}\)\nonumber\\
\leq& P_kP_{k-1}^{-1}\mathbb{E}\tilde\theta_{k-1}\tilde \theta_{k-1}^TP_{k-1}^{-1}P_k+O\({1}/{k^{2}}\)\nonumber\\
=& {P}_{k}{P}_{0}^{-1}\mathbb{E}\tilde\theta_{0}
\tilde\theta_{0}^T{P}_{0}^{-1}{P}_{k}
+O\(\sum_{l=1}^{k}{P}_{k}{P}_{l}^{-1}\frac{1}{l^2}{P}_{l}^{-1}{P}_{k}\)\nonumber\\
=&O\({1}/{k}\).\nonumber
\end{align}
Therefore, $\mathbb{E}\|\tilde\theta_k\|^2=\tr(\mathbb{E}\tilde\theta_k \tilde\theta_k^T)=O\(\frac{1}{k}\)$.\qed
\vspace{-5pt}
\begin{remark}
The key difficulty of this proof is that the weight coefficients $\alpha_{i,k}$ and $\beta_k$ are related to the estimates, which means the weight coefficients are stochastic and coupled with estimation errors.
In addition, to construct an optimal algorithm and achieve that $P_k$ becomes close to the CR lower bound, the weight coefficients cannot be arbitrarily adjusted like the WQNP algorithm.
Therefore, we try to achieve the optimal convergence rate by introducing high-order moments and the inner product of the estimation errors following \cite{ZWZ2021}. However, this method can only reach the convergence rate as $O\(\frac{\log k}{k}\)$ due to the loss of matrix scaling.
In the inner product type Lyapunov function method, it is inevitable to use the technique of matrix scaling to constant coefficients, and the loss caused by matrix scaling makes it impossible to achieve the optimal convergence rate. In order to avoid it, a feasible method is using the cross product type Lyapunov function, which is the cross product of the estimation errors.
However, the projection operator makes us unable to directly iterate the cross product of estimation errors.
 Therefore, we calculate the difference between the projection and no-projection algorithms by using higher moments of estimation errors and Markov inequality, and then, estimate the cross product of estimation errors with the no-projection algorithm.
\end{remark}

\subsection{Proof of Theorem \ref{thm:pktocr}}\label{proof:thm:pktocr}
By Assumption \ref{AI} and (\ref{crlb}),
$\Delta_k=\(\sum_{l=1}^k\rho_l\phi_l\phi_l^T\)^{-1}
=O\(\frac{1}{k}\)$.
Denote
$\beta(x)=\sum_{i=1}^{m+1}\frac{\(f(C_i-x)-f(C_{i-1}-x)\)^2} {F(C_i-x)-F(C_{i-1}-x)}$.
Then, ${\rho}_k={\beta}(\phi_{k}^T\theta)$ and $\hat{\beta}_k={\beta}(\phi_{k}^T\hat{\theta}_{k-1})$. By Assumption \ref{AI} and the continuity of $f(x)$ and $F(x)$, there exists $\hat{\zeta}_{k}$ that is between $\phi^T_ {k}\hat\theta_{k-1}$ and $\phi^T_{k}\hat \theta_k$ such that
\vspace{-8pt}
\begin{align}\label{beta_pk}
\left|\hat\beta_k-\rho_k\right|
&=\left|\beta(\phi_{k}^T\hat{\theta}_{k-1})-\beta(\phi_{k}^T\theta)\right|\nonumber\\
&=\left|\beta^{'}(\hat{\zeta}_k)\phi_{k}^T\tilde{\theta}_{k-1})\right|
=O\(\|\tilde\theta_{k-1}\|\).
\end{align}
{From Theorem \ref{thm:opt-cov}, we have $\mathbb{E}\|\tilde\theta_{l-1}\|\leq\sqrt{\mathbb{E}\|\tilde\theta_{l-1}\|^2} =O\(\frac{1}{\sqrt{l}}\)$. Noticing $\|\phi_k\|\leq \bar\phi$ and $\Delta_k=O\({1}/{k}\)$, we have $\Delta_k^{\frac{1}{2}}\hat{P}_k\Delta_k^{\frac{1}{2}}=o(1)$ and $\Delta_k^{\frac{1}{2}}\sum_{l=1}^k O\(\mathbb{E}\|\tilde\theta_{l-1}\|\) \phi_{l}\phi_{l}^{T}\Delta_k^{\frac{1}{2}}=o(1)$. And then, by $\rho_k=\beta_k$} and (\ref{beta_pk}),
\vspace{-10pt}
\begin{align}
& \mathbb{E}k \hat{P}_k
=\mathbb{E}k \(\Delta_k^{-1}+\sum_{l=1}^k {(\hat\beta_l-\beta_l)\phi_{l}\phi_{l}^{T}}+\hat{P}_0\)^{-1}\nonumber\\
=&\mathbb{E}k \Delta_k^{\frac{1}{2}}\(I +{\Delta_k^{\frac{1}{2}}\sum_{l=1}^k O\(\|\tilde\theta_{l-1}\|\)\phi_{l}\phi_{l}^{T}\Delta_k^{\frac{1}{2}}
+\Delta_k^{\frac{1}{2}}\hat{P}_0\Delta_k^{\frac{1}{2}}} \)^{-1} \Delta_k^{\frac{1}{2}}\nonumber\\
=&{\mathbb{E}k \Delta_k^{\frac{1}{2}}\left(I +\Delta_k^{\frac{1}{2}}\sum_{l=1}^k O\(\|\tilde\theta_{l-1}\|\)\phi_{l}\phi_{l}^{T}\Delta_k^{\frac{1}{2}}
+\Delta_k^{\frac{1}{2}}\hat{P}_0\Delta_k^{\frac{1}{2}}\right.}\nonumber\\
&{\left.+\sum_{i=2}^{\infty}\(\Delta_k^{\frac{1}{2}}\sum_{l=1}^k O\(\|\tilde\theta_{l-1}\|\)\phi_{l}\phi_{l}^{T}\Delta_k^{\frac{1}{2}}
+\Delta_k^{\frac{1}{2}}\hat{P}_0\Delta_k^{\frac{1}{2}}\)^i\right) \Delta_k^{\frac{1}{2}}}\nonumber\\
=&k\Delta_k + O\( k \Delta_k \sum_{l=1}^k O\(\mathbb{E}\|\tilde\theta_{l-1}\|\)
\phi_{l}\phi_{l}^{T}\Delta_k +k\Delta_k\hat{P}_0\Delta_k \)\nonumber\\
=&k\Delta_k +o(1)=k\Delta_k,\nonumber
\end{align}
where the fifth equality is got by Taylor expansion of the symmetric matrix, i.e., $(I+A)^{-1}=I+\sum^{\infty}_{k=1}(-1)^kA^k$ for the symmetric matrix $A$, and the sixth equality is got by Lyapunov inequality.\\
Therefore, $\lim_{k\to\infty}k(\mathbb{E}\hat{P}_k-\Delta_k)=0$.\qed

\subsection{Proof of Theorem \ref{thm:ae}}\label{proof:thm:ae}
Based on Theorem \ref{thm:opt-cov} and (\ref{wj3}), we have
\vspace{-8pt}
\begin{align}
\mathbb{E}\tilde\theta_k\tilde\theta_k^T
\leq& \mathbb{E}\tilde\theta_{k-1}\tilde\theta_{k-1}^T
-\mathbb{E}\tilde\theta_{k-1}\tilde\theta_{k-1}^T{a}_k\beta_k\phi_{k}\phi_{k}^T{P}_{k-1}
\nonumber\\
&-\mathbb{E}{a}_k{P}_{k-1}\beta_k\phi_{k}\phi_{k}^T\tilde\theta_{k-1}\tilde\theta_{k-1}^T
+\mathbb{E}\tilde\theta_{k-1}\tilde\theta_{k-1}^T\phi_{k}\phi_{k}^T\nonumber\\
&\quad\cdot\(\beta_{k}{a}_kP_{k-1}
-\sum_{i=1}^{m} \(\hat\alpha_{i+1,k}-\hat\alpha_{i,k}\)
\grave{f}_{i,k}\hat{a}_k\hat{P}_{k-1}\)\nonumber\\
&+\mathbb{E}\(\beta_{k}{a}_kP_{k-1}
-\sum_{i=1}^{m} \(\hat\alpha_{i+1,k}-\hat\alpha_{i,k}\)
\grave{f}_{i,k}\hat{a}_k\hat{P}_{k-1}\)\nonumber
\end{align}
\vspace{-8pt}
\begin{align}\label{ae1}
&\quad\cdot\phi_{k}\phi_{k}^T\tilde\theta_{k-1}\tilde\theta_{k-1}^T+\mathbb{E}\sum_{i=1}^{m+1}{H}_{i,k}
\(\hat\alpha_{i,k}^2\hat{a}_k^2\hat{P}_{k-1}\phi_{k}\right.\nonumber\\
&\left.\quad\cdot\phi_{k}^T\hat{P}_{k-1}-\alpha_{i,k}^2{P}_{k-1}\phi_{k}\phi_{k}^T{P}_{k-1}\)\nonumber\\
&+\sum_{i=1}^{m+1}\frac{{h}_{i,k}^2}{{H}_{i,k}}
{P}_{k-1}\phi_{k}\phi_{k}^T{P}_{k-1}+o\(\frac{1}{k^{2}}\),
\end{align}
where $P_k$, ${\beta} _k$, $\alpha_{i,k}$ and $a_k$ are defined in the proof of Theorem \ref{thm:opt-cov}.  Then, from
(\ref{hatP}), (\ref{Pk}) and (\ref{wj31}), we have
\vspace{-8pt}
\begin{align}
&\sum_{i=1}^{m+1}{H}_{i,k}
\(\hat\alpha_{i,k}^2\hat{a}_k^2\hat{P}_{k-1}\phi_{k}\phi_{k}^T\hat{P}_{k-1}
-\alpha_{i,k}^2{P}_{k-1}\phi_{k}\phi_{k}^T{P}_{k-1}\)\nonumber\\
=&O\(\frac{1}{k^2}\)\sum_{i=1}^{m+1} |\hat{\alpha}_{i,k}-{\alpha}_{i,k}|
=O\(\frac{1}{k^2}\)\sum_{i=1}^{m+1}| {\alpha}_i^{'}(\hat{\xi}_{i,k})\phi_k^T\tilde\theta_k|
\nonumber\\
=&O\({1}/{k^2}\)\cdot\|\tilde\theta_k\|.\nonumber
\end{align}
where $\hat{\xi}_{i,k}$ is between $\phi^T_ {k}\theta$ and $\phi^T_{k}\hat\theta_{k-1}$. From Theorem \ref{thm:opt-cov} and $\mathbb{E}\|\tilde\theta_k\|\leq\sqrt{\mathbb{E}\|\tilde\theta_k\|^2}
= O\(\frac{1}{\sqrt{k}}\)$, we have
\vspace{-10pt}
\begin{align}\label{ae11}
&\mathbb{E}\sum_{i=1}^{m+1}{H}_{i,k} \(\hat\alpha_{i,k}^2\hat{a}_k^2\hat{P}_{k-1}\phi_{k}\phi_{k}^T\hat{P}_{k-1}
-\alpha_{i,k}^2{P}_{k-1}\phi_{k}\phi_{k}^T{P}_{k-1}\)\nonumber\\
=&O\({1}/{k^2}\)\cdot\mathbb{E}\|\tilde\theta_k\|
=o\({1}/{k^2}\).
\end{align}
Next, we will show $P_{k-1}-P_{k}=O\(\frac{1}{k^2}\)$. Noticing $P_ {k}=P_{k-1}-a_k\beta_kP_{k-1}\phi_{k}\phi_{k}^TP_{k-1}$, where $\beta_k=\beta(\phi_ {k}^T\theta)$ is bounded and positive, we have
\vspace{-8pt}
\begin{align}\label{PkPl}
\|P_{k-1}-P_{k}\|
\leq \beta_k\bar\phi^2\|P_{k-1}\|^2=O\({1}/{k^2}\).
\end{align}
From Theorem  \ref{thm:opt-cov}, substituting (\ref{wj4}), (\ref{ae11}) and (\ref{PkPl}) into (\ref{ae1}) gives
\vspace{-8pt}
\begin{align}
&\mathbb{E}\tilde\theta_k\tilde\theta_k^T
= \mathbb{E}\tilde\theta_{k-1}\tilde\theta_{k-1}^T
-\mathbb{E}\tilde\theta_{k-1}\tilde\theta_{k-1}^Ta_k\beta_k\phi_{k}\phi_{k}^T{P}_{k-1}\nonumber\\ &\qquad\qquad-\mathbb{E}a_k{P}_{k-1}\beta_k\phi_{k}\phi_{k}^T\tilde\theta_{k-1}\tilde\theta_{k-1}^T\nonumber\\
&\qquad\qquad+\sum_{i=1}^{m+1}\frac{{h}_{i,k}^2}{{H}_{i,k}}{P}_{k}\phi_{k}\phi_{k}^T{P}_{k}
+o\(\frac{1}{k^2}\)\nonumber\\
=& {P}_{k}{P}_{k-1}^{-1}\mathbb{E}\tilde\theta_{k-1}
\tilde\theta_{k-1}^T{P}_{k-1}^{-1}{P}_{k}
+\sum_{i=1}^{m+1}\frac{{h}_{i,k}^2}{{H}_{i,k}}{P}_{k}\phi_{k}\phi_{k}^T{P}_{k}
+o\(\frac{1}{k^2}\)\nonumber\\
=& {P}_{k}{P}_{0}^{-1}\mathbb{E}\tilde\theta_{0}
\tilde\theta_{0}^T{P}_{0}^{-1}{P}_{k}
+o\(\sum_{l=1}^k{P}_{k}{P}_{l}^{-1}\frac{1}{l^2}
{P}_{l}^{-1}{P}_{k}\)\nonumber\\
&+\sum_{l=1}^k\sum_{i=1}^{m+1}\frac{{h}_{i,l}^2}{{H}_{i,l}}{P}_{k}{P}_{l}^{-1}{P}_{l}
\phi_{l}\phi_{l}^T{P}_{l}{P}_{l}^{-1}{P}_{k}\nonumber\\
=& O\(\frac{1}{k^2}\)+{P}_{k}\Delta_k^{-1}{P}_{k}
+o\(\sum_{l=1}^{k}{P}_{k}{P}_{l}^{-1}\frac{1}{l^2}{P}_{l}^{-1}{P}_{k}\)\nonumber\\
=& o\({1}/{k}\)+{P}_{k}\Delta_k^{-1}{P}_{k} =\Delta_k+o\({1}/{k}\),\nonumber
\end{align}
which implies the conclusion.\qed

\section{Numerical example}\label{sec:sim}
\begin{example}\label{ex1}
Consider a third-order system
\begin{gather}\label{ex1_m}
y_k=a_1+a_2u_{k}+a_3u_{k-1}+d_k, k\geq 1,
\end{gather}
where the parameter $\theta=[a_1, a_2, a_3]^T=[-0.5,1,-1]^T$ is unknown, and the prior information is $\theta\in\Omega=[-3,3]\times[0,2]\times[-2,0]$; the system noise $d_{k}$ follows {$N(0,1.5^2)$}; the output $y_{k}$ only can be measured by quantized sensor $q_k=\sum_{i=0}^{3} i I_{\{C_{i} < y_k \leq C_{i+1}\}}$ with $C_1=-1,C_2=0,C_3=0.5$.
  and the regressor $\phi_k=[1, u_k, u_{k-1}]^T$ is generated by $u_{3l}=-2+e_{3l}$, $u_{3l+1}=e_{3l+1}$ and $u_{3l+2}=0.5+e_{3l+2}$ for $l=0, 1, 2, \ldots$, where $e_k$ is randomly chosen in the interval $[0, 0.1]$. In this case, $\phi_k$ follows the conditions of Assumption \ref{AI}.

\begin{figure}[htbp]
	\centering
	\includegraphics[width=7.4cm]{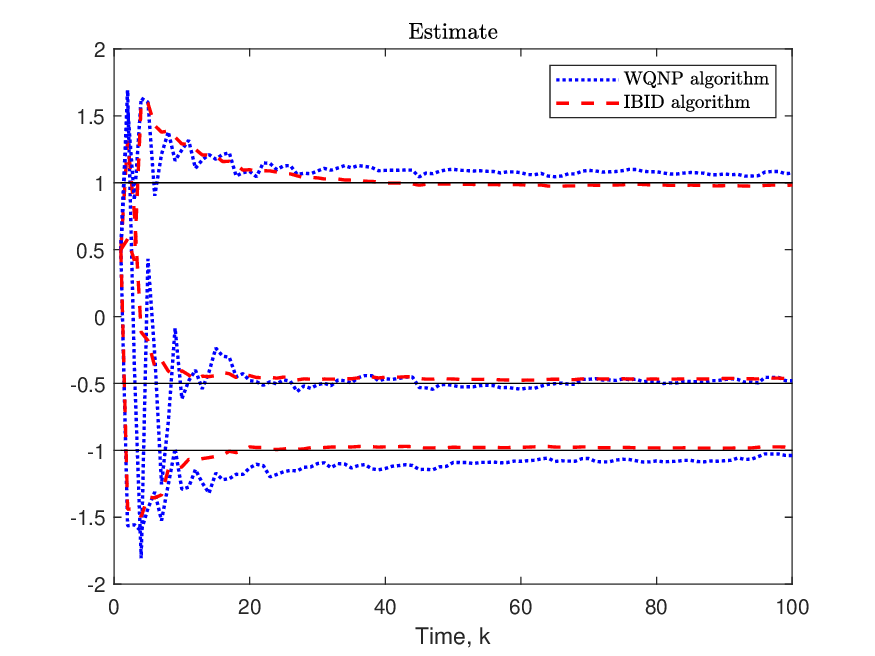}
	\caption{Convergence of the WQNP/IBID algorithm.} %
	\label{figc}
\end{figure}

\begin{figure}[htbp]
	\centering
	\includegraphics[width=7.4cm]{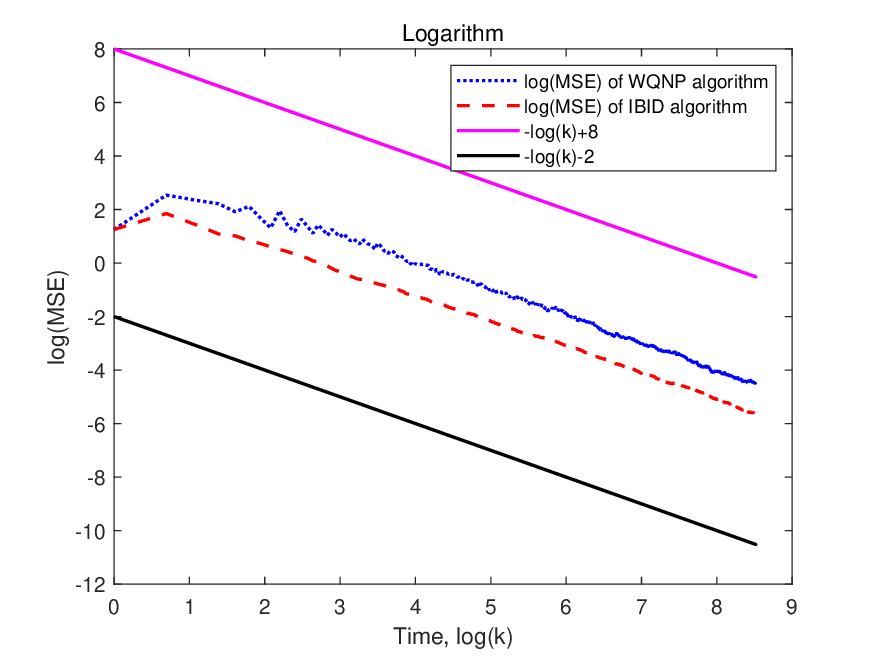}
	\caption{Convergence rate of the WQNP/IBID algorithm.} %
	\label{figcr}
\end{figure}

We apply the WQNP algorithm and the IBID algorithm to give the estimates, where the weight coefficients of the WQNP algorithm are $[\alpha_{1,k},\alpha_{2,k},\alpha_{3,k},\alpha_{4,k},\beta_k]=[1,8,14,20,0.5]$, and ones of the IBID algorithm are given in (\ref{opt:ab}). Here we repeat the simulation 500 times with the same initial estimate $\hat{\theta}_0=[1/2,1/2,1/2]^T$ and $\hat{P}_0=P_0=3I_3$ to establish the empirical variance of the estimation errors representing the mean square errors.

\begin{figure}[htbp]
	\centering
	\includegraphics[width=7.6cm]{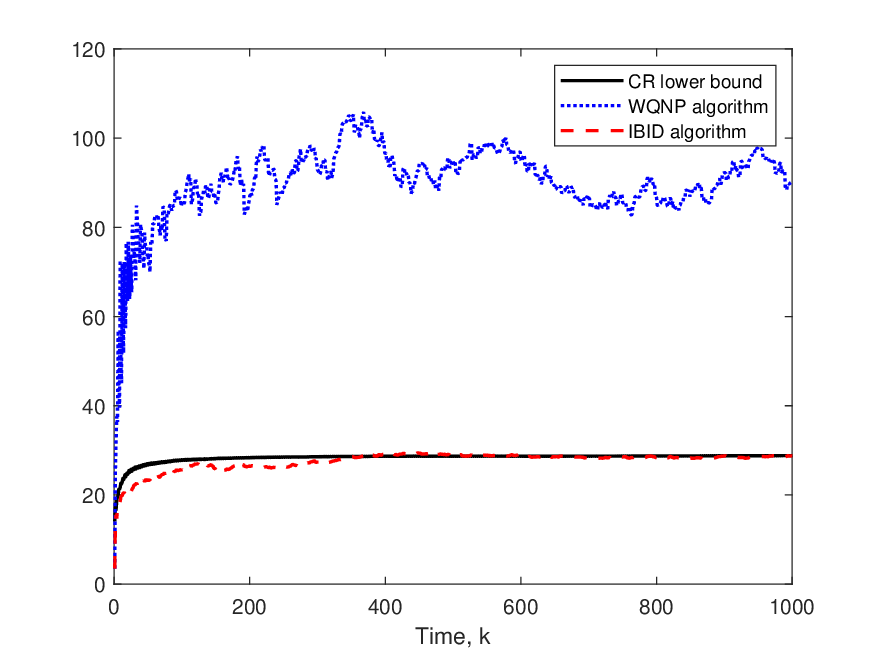}
	\caption{Comparison between the empirical variance ($k\tilde{\theta}_k^T\tilde{\theta}_k$) of the WQNP/IBID algorithm and CR lower bound ($k\tr(\Delta_k)$).} %
	\label{figcrlb}
\end{figure}

The simulation results are given in Figures \ref{figc}-\ref{figcrlb}. Figure \ref{figc} shows the convergence results of the WQNP algorithm with constant weights and the IBID algorithm, and it can be seen that the IBID algorithm converges to the true parameter faster than the WQNP algorithm. From Figure \ref{figcr}, we see that the logarithm of the mean square error (MSE) is linear with the logarithm of the index $k$, which indicates that the mean square convergence rates of the WQNP algorithm and the IBID algorithm are $O(\frac{1}{k})$. Besides, Figure \ref{figcrlb} shows that the comparison results between the empirical variance of estimation errors of the two algorithms and the CR lower bound.
It is obvious that the IBID algorithm performs better than the WQNP algorithm and can reach asymptotically the CR lower bound, meaning the IBID algorithm is asymptotically efficient.
\end{example}

\section{Concluding remarks}\label{sec:con}

{This paper focuses on how to design an optimal identification algorithm under quantized observations.}
First, a weighted Quasi-Newton type projection algorithm is proposed to identify dynamical systems with quantized observations under bounded persistent excitations. Then, based on the adaptive design on the weight coefficients of the WQNP algorithm via the structure of CR lower bound, an IBID algorithm is constructed. And the mean square convergence rate of the algorithm can reach the reciprocal of the number of observations.
 Moreover, the asymptotic efficiency of the IBID algorithm is established, which means its optimality.

{These optimality results lay a foundation for designing appropriate communication protocol (threshold choice) and communication times to achieve the best identification performance under same communication resources.} Correspondingly, future work is directed at studying sensor threshold selection to achieve optimal utility of communication bandwidth resources in enhancing identification accuracy.


\begin{thebibliography}{36}
\expandafter\ifx\csname natexlab\endcsname\relax\def\natexlab#1{#1}\fi
\providecommand{\url}[1]{\texttt{#1}}
\providecommand{\href}[2]{#2}
\providecommand{\path}[1]{#1}
\providecommand{\DOIprefix}{doi:}
\providecommand{\ArXivprefix}{arXiv:}
\providecommand{\URLprefix}{URL: }
\providecommand{\Pubmedprefix}{pmid:}
\providecommand{\doi}[1]{\href{http://dx.doi.org/#1}{\path{#1}}}
\providecommand{\Pubmed}[1]{\href{pmid:#1}{\path{#1}}}
\providecommand{\bibinfo}[2]{#2}
\ifx\xfnm\relax \def\xfnm[#1]{\unskip,\space#1}\fi
\bibitem[{Auber et~al.(2018)Auber, Pouliquen, Pigeon, Chapon and
  Moussay}]{APPCM2018}
\bibinfo{author}{Auber, R.}, \bibinfo{author}{Pouliquen, M.},
  \bibinfo{author}{Pigeon, E.}, \bibinfo{author}{Chapon, P. A.},
  \bibinfo{author}{Moussay, S.}, \bibinfo{year}{(2018)}.
\newblock \bibinfo{title}{Activity recognition from binary data}, in:
  \bibinfo{booktitle}{2018 UKACC 12th International Conference on Control}, pp.
  \bibinfo{pages}{158--162}.
\bibitem[{Calamai and Mor\'{e}(1987)}]{CM1987}
\bibinfo{author}{Calamai, P. H.}, \bibinfo{author}{Mor\'{e}, J. J.},
  \bibinfo{year}{(1987)}.
\newblock \bibinfo{title}{Projected gradient methods for linearly constrained
  problems}.
\newblock \bibinfo{journal}{Mathematical Programming}, \bibinfo{volume}{39},
  \bibinfo{pages}{93--116}.
\bibitem[{{Carbone} et~al.(2020){Carbone}, {Schoukens} and
  {Moschitta}}]{CSM2020}
\bibinfo{author}{{Carbone}, P.}, \bibinfo{author}{{Schoukens}, J.},
  \bibinfo{author}{{Moschitta}, A.}, \bibinfo{year}{(2020)}.
\newblock \bibinfo{title}{Quick estimation of periodic signal parameters from
  1-bit measurements},
\newblock \bibinfo{journal}{IEEE Transactions on Instrumentation and
  Measurement}, \bibinfo{volume}{69}, \bibinfo{pages}{339--353}.
\bibitem[{Casini et~al.(2011)Casini, Garulli and Vicino}]{CGV2011}
\bibinfo{author}{Casini, M.}, \bibinfo{author}{Garulli, A.},
  \bibinfo{author}{Vicino, A.}, \bibinfo{year}{(2011)}.
\newblock \bibinfo{title}{Input design in worst-case system identification
  using binary sensors}.
\newblock \bibinfo{journal}{IEEE Transactions on Automatic Control},
  \bibinfo{volume}{56}, \bibinfo{pages}{1186--1191}.
\bibitem[{Chen(2002)}]{Chen2002}
\bibinfo{author}{Chen, H. F.}, \bibinfo{year}{(2002)}.
\newblock \bibinfo{title}{Stochastic Approximation and Its Applications}.
\newblock \bibinfo{publisher}{Kluwer Academic Publishers, Dordrecht}.
\bibitem[{Cs\'{a}ji and Weyer(2012)}]{CW2012}
\bibinfo{author}{Cs\'{a}ji, B. C.}, \bibinfo{author}{Weyer, E.},
  \bibinfo{year}{(2012)}.
\newblock \bibinfo{title}{Recursive estimation of {ARX} systems using binary
  sensors with adjustable thresholds}.
\newblock \bibinfo{journal}{IFAC Proceedings Volumes}, \bibinfo{volume}{45},
  \bibinfo{pages}{1185--1190}.
\bibitem[{Dargie and Poellabauer(2010)}]{DC2010}
\bibinfo{author}{Dargie, W.}, \bibinfo{author}{Poellabauer, C.},
  \bibinfo{year}{(2010)}.
\newblock \bibinfo{title}{Fundamentals of wireless sensor networks: theory and
  practice}.
\newblock \bibinfo{publisher}{Hoboken, NJ, USA: Wiley}.
\bibitem[{Gagliardi et~al.(2021)Gagliardi, Mari, Tedesco and
  Casavola}]{GMTC2021}
\bibinfo{author}{Gagliardi, G.}, \bibinfo{author}{Mari, D.},
  \bibinfo{author}{Tedesco, F.}, \bibinfo{author}{Casavola, A.},
  \bibinfo{year}{(2021)}.
\newblock \bibinfo{title}{An {Air-to-Fuel} ratio estimation strategy for
  turbocharged spark-ignition engines based on sparse binary hego sensor
  measures and hybrid linear observers}.
\newblock \bibinfo{journal}{Control Engineering Practice},
  \bibinfo{volume}{107}, \bibinfo{pages}{104694}.
\bibitem[{Ghysen(2003)}]{G2003}
\bibinfo{author}{Ghysen, A.}, \bibinfo{year}{(2003)}.
\newblock \bibinfo{title}{The origin and evolution of the nervous system}.
\newblock \bibinfo{journal}{Control Engineering Practice}, \bibinfo{volume}{47},
  \bibinfo{pages}{555--562}.
\bibitem[{Godoy et~al.(2011)Godoy, Goodwin, Ag\"{u}ero, Marelli and
  Wigren}]{GGAMW2011}
\bibinfo{author}{Godoy, B.}, \bibinfo{author}{Goodwin, G.},
  \bibinfo{author}{Ag\"{u}ero, J.}, \bibinfo{author}{Marelli, D.},
  \bibinfo{author}{Wigren, T.}, \bibinfo{year}{(2011)}.
\newblock \bibinfo{title}{On identification of {FIR} systems having quantized
  output data}.
\newblock \bibinfo{journal}{Automatica}, \bibinfo{volume}{47},
  \bibinfo{pages}{1905--1915}.
\bibitem[{Guo and Diao(2020)}]{GD2020}
\bibinfo{author}{Guo, J.}, \bibinfo{author}{Diao, J.}, \bibinfo{year}{(2020)}.
\newblock \bibinfo{title}{Prediction-based event-triggered identification of
  quantized input {FIR} systems with quantized output observations}.
\newblock \bibinfo{journal}{Science China Information Sciences},
  \bibinfo{volume}{63}, \bibinfo{pages}{112201}.
\bibitem[{Guo et~al.(2015)Guo, Wang, Yin, Zhao and Zhang}]{GWYZZ2015}
\bibinfo{author}{Guo, J.}, \bibinfo{author}{Wang, L. Y.}, \bibinfo{author}{Yin,
  G.}, \bibinfo{author}{Zhao, Y. L.}, \bibinfo{author}{Zhang, J. F.},
  \bibinfo{year}{(2015)}.
\newblock \bibinfo{title}{Asymptotically efficient identification of {FIR}
  systems with quantized observations and general quantized inputs}.
\newblock \bibinfo{journal}{Automatica}, \bibinfo{volume}{57},
  \bibinfo{pages}{113--122}.
\bibitem[{Guo and Zhao(2013)}]{GZ2013}
\bibinfo{author}{Guo, J.}, \bibinfo{author}{Zhao, Y. L.}, \bibinfo{year}{(2013)}.
\newblock \bibinfo{title}{Recursive projection algorithm on {FIR} system
  identification with binary-valued observations}.
\newblock \bibinfo{journal}{Automatica}, \bibinfo{volume}{49},
  \bibinfo{pages}{3396--3401}.
\bibitem[{Guo and Zhao(2014)}]{GZ2014}
\bibinfo{author}{Guo, J.}, \bibinfo{author}{Zhao, Y. L.}, \bibinfo{year}{(2014)}.
\newblock \bibinfo{title}{Identification of the gain system with quantized
  observations and bounded persistent excitations}.
\newblock \bibinfo{journal}{Science China Information Sciences},
  \bibinfo{volume}{57}, \bibinfo{pages}{012205}.
\bibitem[{Guo(2020)}]{S_Guo2020}
\bibinfo{author}{Guo, L.}, \bibinfo{year}{(2020)}.
\newblock \bibinfo{title}{Time-Varying Stochastic Systems, Stability and
  Adaptive Theory, Second Edition}.
\newblock \bibinfo{publisher}{Science Press, Beijing}.
\bibitem[{Gustafsson and Karlsson(2009)}]{GK2009}
\bibinfo{author}{Gustafsson, F.}, \bibinfo{author}{Karlsson, R.},
  \bibinfo{year}{(2009)}.
\newblock \bibinfo{title}{Statistical results for system identification based
  on quantized observations}.
\newblock \bibinfo{journal}{Automatica}, \bibinfo{volume}{45},
  \bibinfo{pages}{2794--2801}.

\bibitem[{Ljung and S\"{o}derstr\"{o}m (1983)}]{S_LS1983}
\bibinfo{author}{Ljung, L.}, \bibinfo{author}{S\"{o}derstr\"{o}m, T.,}, \bibinfo{year}{(1983)}.
\newblock \bibinfo{title}{Theory and Practice of Recursive Identification}.
\newblock \bibinfo{publisher}{The MIT Press, London}.
\bibitem[{K.~Sohraby and Znati(2007)}]{SMZ2007}
\bibinfo{author}{Sohraby, D. M. K.}, \bibinfo{author}{Znati, T.},
  \bibinfo{year}{(2007)}.
\newblock \bibinfo{title}{Wireless sensor networks: technology, protocols, and
  applications}.
\newblock \bibinfo{publisher}{John Wiley and Sons}.
\bibitem[{Risuleo et~al.(2020)Risuleo, Bottegal and Hjalmarsson}]{RBH2020}
\bibinfo{author}{Risuleo, R. S.}, \bibinfo{author}{Bottegal, G.},
  \bibinfo{author}{Hjalmarsson, H.}, \bibinfo{year}{(2020)}.
\newblock \bibinfo{title}{Identification of linear models from quantized data:
  A midpoint-projection approach}.
\newblock \bibinfo{journal}{IEEE Transactions on Automatic Control},
  \bibinfo{volume}{65}, \bibinfo{pages}{2801--2813}.
\bibitem[{Song(2018)}]{S2018}
\bibinfo{author}{Song, Q.}, \bibinfo{year}{(2018)}.
\newblock \bibinfo{title}{Recursive identification of systems with
  binary-valued outputs and with {ARMA} noises}.
\newblock \bibinfo{journal}{Automatica}, \bibinfo{volume}{93},
  \bibinfo{pages}{106--113}.
\bibitem[{Wang et~al.(2002)Wang, Kim and Sun}]{WKS2002}
\bibinfo{author}{Wang, L. Y.}, \bibinfo{author}{Kim, Y. W.},
  \bibinfo{author}{Sun, J.}, \bibinfo{year}{(2002)}.
\newblock \bibinfo{title}{Prediction of oxygen storage capacity and stored
  {NO$_x$} by {HEGO} sensors for improved lnt control strategies}, in:
  \bibinfo{booktitle}{Dynamic Systems and Control of ASME International
  Mechanical Engineering Congress and Exposition}, pp.
  \bibinfo{pages}{777--785}.
\bibitem[{Wang et~al.(2010)Wang, Yin, Zhang and Zhao}]{WYZZ2010}
\bibinfo{author}{Wang, L. Y.}, \bibinfo{author}{Yin, G.},
  \bibinfo{author}{Zhang, J. F.}, \bibinfo{author}{Zhao, Y.},
  \bibinfo{year}{(2010)}.
\newblock \bibinfo{title}{System Identification with Quantized Observations}.
\newblock \bibinfo{publisher}{Birkhauser Boston}.
\bibitem[{Wang and Yin(2007)}]{WangandYin2007}
\bibinfo{author}{Wang, L. Y.}, \bibinfo{author}{Yin, G.},
  \bibinfo{year}{(2007)}.
\newblock \bibinfo{title}{Asymptotically efficient parameter estimation using
  quantized output observations}.
\newblock \bibinfo{journal}{Automatica}, \bibinfo{volume}{43},
  \bibinfo{pages}{1178--1191}.
\bibitem[{Wang et~al.(2003)Wang, Zhang and Yin}]{WZY2003}
\bibinfo{author}{Wang, L. Y.}, \bibinfo{author}{Zhang, J. F.},
  \bibinfo{author}{Yin, G.}, \bibinfo{year}{(2003)}.
\newblock \bibinfo{title}{System identification using binary sensors}.
\newblock \bibinfo{journal}{IEEE Transactions on Automatic Control},
  \bibinfo{volume}{48}, \bibinfo{pages}{1892--1907}.
\bibitem[{Wang et~al.(2018)Wang, Tan and Zhao}]{WTZ2018}
\bibinfo{author}{Wang, T.}, \bibinfo{author}{Tan, J.}, \bibinfo{author}{Zhao,
  Y. L.}, \bibinfo{year}{(2018)}.
\newblock \bibinfo{title}{Asymptotically efficient non-truncated identification
  for {FIR} systems with binary-valued outputs}.
\newblock \bibinfo{journal}{Science China Information Sciences},
  \bibinfo{volume}{61}, \bibinfo{pages}{129208}.
\bibitem[{Wang et~al.(2022)Wang, Zhao, Zhang and Guo}]{WZZG2022}
\bibinfo{author}{Wang, Y.}, \bibinfo{author}{Zhao, Y. L.}, \bibinfo{author}{Zhang,
  J.F.}, \bibinfo{author}{Guo, J.}, \bibinfo{year}{(2022)}.
\newblock \bibinfo{title}{A unified identification algorithm of {FIR} systems
  based on binary observations with time-varying thresholds}.
\newblock \bibinfo{journal}{Automatica}, \bibinfo{volume}{135},
  \bibinfo{pages}{109990}.
\bibitem[{Wu et~al.(2013)Wu, Wang and Ye}]{WWY2013}
\bibinfo{author}{Wu, H.}, \bibinfo{author}{Wang, W.}, \bibinfo{author}{Ye, H.},
  \bibinfo{year}{2013}.
\newblock \bibinfo{title}{Lower bounds in parameter estimation based on
  quantized measurements}, in: \bibinfo{booktitle}{52nd IEEE Conference on
  Decision and Control}, pp. \bibinfo{pages}{6341--6346}.
\bibitem[{Yang and Fang(2014)}]{YF2014}
\bibinfo{author}{Yang, X.}, \bibinfo{author}{Fang, H. T.}, \bibinfo{year}{(2014)}.
\newblock \bibinfo{title}{Asymptotically efficient recursive identification
  method for {FIR} system with quantized observations}, in:
  \bibinfo{booktitle}{Proceedings of the 33rd Chinese Control Conference}, pp.
  \bibinfo{pages}{6832--6837}.
\bibitem[{You(2015)}]{You2015}
\bibinfo{author}{You, K.}, \bibinfo{year}{(2015)}.
\newblock \bibinfo{title}{Recursive algorithms for parameter estimation with
  adaptive quantizer}.
\newblock \bibinfo{journal}{Automatica}, \bibinfo{volume}{52},
  \bibinfo{pages}{192--201}.
\bibitem[{Zhang et~al.(2021)Zhang, Wang and Zhao}]{ZWZ2021}
\bibinfo{author}{Zhang, H.}, \bibinfo{author}{Wang, T.}, \bibinfo{author}{Zhao,
  Y. L.}, \bibinfo{year}{(2021)}.
\newblock \bibinfo{title}{Asymptotically efficient recursive identification of
  {FIR} systems with binary-valued observations}.
\newblock \bibinfo{journal}{IEEE Transactions on Systems, Man, and Cybernetics:
  Systems}, \bibinfo{volume}{51}, \bibinfo{pages}{2687--2700}.
\bibitem[{Zhang et~al.(2022)Zhang, Zhao and Guo}]{ZZG2022}
\bibinfo{author}{Zhang, L.}, \bibinfo{author}{Zhao, Y. L.}, \bibinfo{author}{Guo,
  L.}, \bibinfo{year}{(2022)}.
\newblock \bibinfo{title}{Identification and adaptation with binary-valued
  observations under non-persistent excitation condition}.
\newblock \bibinfo{journal}{Automatica}, \bibinfo{volume}{138},
  \bibinfo{pages}{110158}.
\bibitem[{Zhao et~al.(2023)Zhao, Zhang, Wang and Kang}]{ZZWK2023}
 \bibinfo{author}{Zhao, Y. L.}, \bibinfo{author}{Zhang, H.},
 \bibinfo{author}{Wang, T.}, \bibinfo{author}{Kang, G.},
 \bibinfo{year}{(2023)}.
\newblock \bibinfo{title}{System identification under saturated precise or set-valued measurements}.
\newblock \bibinfo{journal}{Science China Information Sciences},
\bibinfo{volume}{66},  \bibinfo{pages}{112204}.
\end{thebibliography}

\end{document}